\documentclass[american]{revtex4-2}
\usepackage[T1]{fontenc}
\usepackage[utf8]{inputenc}
\usepackage{float}
\usepackage{booktabs}
\usepackage{bm}
\usepackage{varwidth}
\usepackage{amsmath}
\usepackage{graphicx}

\makeatletter

\providecommand{\tabularnewline}{\\}
\newenvironment{cellvarwidth}[1][t]
    {\begin{varwidth}[#1]{\linewidth}}
    {\@finalstrut\@arstrutbox\end{varwidth}}

\usepackage{graphicx}
\usepackage{tikz}
\usetikzlibrary{arrows.meta,positioning}
\newcommand{\td}{\widetilde}

\newcommand{\argmaxop}{\operatorname*{arg\,max}}
\newcommand{\vrr}{\vrule width 0pt height 10pt depth 5pt}
\newcommand{\vrt}{\vrule width 0pt height 10pt depth 0pt}
\newcommand{\vrd}{\vrule width 0pt height 0pt depth 5pt}

\usepackage{array}
\usepackage{float}

\makeatother

\usepackage{babel}
\begin{document}
\title{Plasma localization of charged debris}
\author{Bikramjit Joardar and Madhurjya P.\ Bora}
\address{\emph{Physics Department, Gauhati University, Guwahati 781014, India}}
\begin{abstract}
We investigate whether the plasma disturbances generated by a charged
debris object can be used to infer its position without directly sampling
the object or its immediate sheath. A one-dimensional open-boundary
electrostatic particle-in-cell (PIC) framework is used to model a
continuously flowing electron-ion plasma containing an initially uncharged
debris object. The debris charges self-consistently and produces an
ion-ion counter-streaming instability (IICSI), which is sustained
and reaches a statistically stationary state. A matching simulation
without debris provides a controlled background against which debris-induced
changes in different plasma fields are identified. The precursor and
wake are examined through their spatial extent, fluctuation power,
and frequency--wavenumber spectra. Weak-form sparse regression is
then used as a supporting tool to recover reduced fluid and kinetic
residuals and to learn an empirical relation connecting the precursor
and wake disturbance envelopes to the equilibrium plasma flow, debris
charge, and distance from the source. The predictive capability is
examined by successively treating each simulation as an independent
case. For each evaluation, the model is constructed from the remaining
simulations, while the excluded simulation is used only to infer the
debris position from its plasma response. For the idealized cold-ion
regime considered here, the procedure successfully localizes the debris
in transonic and supersonic flows, whereas the subsonic cases remain
unresolved. These results demonstrate that debris-induced plasma disturbances
possess a learnable spatial structure and provide a proof-of-principle
route toward future remote localization using more realistic multidimensional
models; they do not constitute an operational debris-detection scheme. 
\end{abstract}
\maketitle

\section{Introduction}\label{sec:Introduction}

The study of plasma interactions with external objects dates back
almost to the beginning of plasma physics itself. Langmuir’s pioneering
investigations of electrodes immersed in ionized gases established
the fundamental concepts of particle collection, floating potential,
and electrostatic sheath formation \citep{MottSmithLangmuir1926}.
These ideas were subsequently extended to probes, material surfaces,
spacecraft, and charged grains. The interaction of external debris
with an ambient space plasma therefore belongs to this long-standing
class of plasma-surface problems. Objects immersed in a plasma acquire
an electrostatic charge through the collection of ambient electrons
and ions, photoelectron emission, and other surface processes. Their
relative motion through the surrounding plasma can produce spatially
asymmetric responses through upstream precursor disturbances and downstream
wakes, which are themselves of considerable fundamental interest.
However, the rapid growth of the low-Earth-orbit (LEO) space-debris
population over recent decades has added a new and practically important
dimension to this classical problem, motivating investigations of
whether plasma disturbances can provide information about debris too
small to be reliably observed by conventional methods. Spacecraft--plasma
interactions have been studied extensively in this context \citep{Hastings1995}.
Particle simulations and spacecraft observations have further shown
that a charged object in a flowing tenuous plasma can generate a measurable
electrostatic wake and that the wake itself can be used as an indirect
plasma diagnostic \citep{Engwall2006a,Engwall2006b,Miyake2013}. Related
particle-in-cell (PIC) studies of charged grains have demonstrated
that relative plasma flow breaks the symmetry of the charging and
produces characteristic density and potential structures downstream
of the object \citep{Miloch2014}.

Charged objects can also excite nonlinear structures in the surrounding
plasma. Laboratory experiments and numerical studies have reported
upstream precursor solitons, downstream wakes, Mach cones, and dispersive
structures generated by plasma flowing past an electrostatic obstacle
\citep{Jaiswal2016,Tiwari2016}. In the specific context of orbital
debris, forced-Korteweg-de Vries (fKdV) models have suggested that
plasma solitary waves may provide signatures of otherwise difficult-to-detect
sub-centimeter objects \citep{Truitt2020}. More recent PIC simulations
have shown that an energetic charged body can produce trailing wakes,
fore-wake shocks, precursor solitons, and particle trapping, all of
which may serve as signatures of its passage through a plasma \citep{Dharodi2023}.
The nonlinear response resulting from plasma--debris interaction
depends strongly on the sign, magnitude, velocity, and temporal variation
of the external charge. Moving external charge perturbations were
shown to generate pinned nonlinear structures and electrostatic dispersive
shocks \citep{SarkarBora2023}. A negatively charged external perturbation
can accelerate ions toward itself from opposite sides, thereby producing
counter-streaming ion populations. When their relative drift is sufficiently
large, the resulting ion-ion counter-streaming instability (IICSI)
develops through phase-space vortices into pinned solitons and electrostatic
turbulence \citep{DasBora2025}. In a recent work, we have shown that
debris-charge fluctuations can in turn modify the plasma response
and produce polarity-dependent chaos and nonlinear Landau damping
\citep{Joardar2025}. These results suggest that the spatially organized
plasma response to a charged object may contain information about
the object even when its statistically equilibrated charge is dominated
by broadband fluctuations.

We note that the primary strategy of most previous studies has been
to investigate the nonlinear structures formed by plasma--debris
interaction. The corresponding inverse problem -- whether the position
of an unknown object can be inferred from remotely measured plasma
fields -- has received considerably less attention. In particular,
it remains useful to determine whether precursor and wake perturbations
contain a sufficiently coherent, source-centered signature to distinguish
a debris-bearing plasma from the matched background and to localize
the external object without sampling its immediate sheath. Toward
this, Sparse Identification of Nonlinear Dynamics (SINDy) provides
an interpretable framework for addressing this question by selecting
a parsimonious governing equation from a library of candidate physical
terms \citep{Brunton2016}. This method has been extended to spatially
distributed systems through sparse partial-differential-equation discovery
\citep{Rudy2017}. For noisy data, weak or integral formulations are
used to avoid direct point-wise differentiation by transferring derivatives
from the measured fields to smooth test functions, substantially improving
robustness \citep{Messenger2021a,Messenger2021b}. This property is
particularly relevant to particle-in-cell data, for which particle
noise and numerical differentiation can otherwise obscure weak localized
signatures.

In this work, we use a numerical analogue of a continuously flowing
tenuous electron--ion plasma containing external debris, implemented
with an open-boundary configuration of our well-benchmarked one-dimensional
\emph{hybrid}-PIC--MCC code \citep{ChangmaiBora2019,ChangmaiBora2020}.
The model represents an idealized regime motivated by topside-ionospheric
conditions. However, we must note that it is \emph{not} a direct numerical
reproduction of a particular LEO orbit. The work has three connected
parts. First, we examine the limitation of an fKdV description once
the counter-streaming response reaches a fully developed nonlinear
kinetic state. Second, we investigate the precursor and wake regions
and identify their signatures in different plasma fields. Third, we
examine whether these signatures can be used to infer the source position
through weak-form residual analysis and an empirical disturbance model.
The present calculation is intended as a proof-of-principle demonstration
of a possible route to localization, not as an operational debris-detection
strategy. SINDy is used only as a supporting analysis tool, the primary
emphasis is on the plasma response and the spatial information carried
by the precursor and wake.

\section{Regime of interest}\label{sec:Regime-of-interest}

The simulation parameters used in this work define an idealized regime
in which the nonlinear response develops clearly within an accessible
computational time. The LEO values quoted below serve only to illustrate
conditions under which the same mechanism may be energetically plausible.
Ionospheric density, temperature, and composition vary strongly with
altitude, location, local time, and solar activity, and no single
parameter set can represent the LEO plasma as a whole. Nevertheless,
the prevalent oxygen $(\textrm{O}^{+})$ and hydrogen $(\textrm{H}^{+})$
ions provide a useful physical context consistent with standard descriptions
of the terrestrial ionosphere \citep{SchunkNagy2009}.

Quantitative modeling of a particular ionospheric environment would
require a variable $\textrm{O}^{+}/\textrm{H}^{+}$ mixture in three-dimensional
geometry and is beyond the present scope. Since our purpose is to
establish the mechanism and examine the associated inverse problem,
we choose parameters for which a sustained nonlinear saturation regime
develops clearly. Unless stated otherwise, the reported calculations
correspond to a singly charged hydrogen plasma with $T_{e0}=1\,\textrm{eV}$,
$T_{i}=0.01\,\textrm{eV}$, and $m_{i}/m_{e}\simeq1836$. The use
of $\textrm{H}^{+}$ is a deliberate choice as it provides a computationally
economical reference system in which the mechanism can be isolated.
The discussion of $\textrm{O}^{+}$ below is limited to ion-mass scaling
only. Since $\omega_{\textrm{pi}}\propto m^{-1/2}_{i}$, replacing
$\textrm{H}^{+}$ by $\textrm{O}^{+}$ reduces the ion plasma frequency
by approximately a factor of four. Consequently, an $\textrm{O}^{+}$
calculation must be evolved for about four times as many electron-normalized
time units to cover the same interval in $\omega_{\textrm{pi}}t$.
When the time step is constrained by the electron dynamics, this corresponds
approximately to a four-fold increase in computational time, assuming
the same grid, particle number, diagnostics, and required ion-dynamical
duration. The hydrogen calculation is therefore used here as the computationally
economical system, while the oxygen-ion and mixed oxygen-hydrogen
cases are reserved for a future dedicated study.

\subsection{Self-excited IICSI}

In this section, we briefly review the plasma configuration and parameter
regime in which a self-excited IICSI may occur in the context of plasma-debris
interaction. Much of this calculation is based on the original work
done by Das and Bora \citep{DasBora2025}.

We start from the 1D Vlasov equation for the ions 
\begin{equation}
\frac{\partial f_{i}}{\partial t}+v\frac{\partial f_{i}}{\partial x}+\frac{e}{m_{i}}E\frac{\partial f_{i}}{\partial v}=0,
\end{equation}
where $f_{i}$ is the ion velocity distribution, $m_{i}$ is the ion
mass, and $E=-\partial\phi/\partial x$ is the electric field with
$\phi$ being the plasma potential. The electrons are assumed to be
Boltzmannian. Assuming a perturbation of the form $\sim e^{-i\omega t+ikx}$,
the linear electrostatic dispersion relation can be written as 
\begin{equation}
1+\frac{1}{k^{2}\lambda^{2}_{De}}-\frac{\omega^{2}_{\textrm{pi}}}{k^{2}n_{0}}\int^{+\infty}_{-\infty}\frac{\partial f_{i0}/\partial v}{v-\omega/k}\,dv=0,\label{eq:iicsi-dispersion}
\end{equation}
where the velocity integral represents the usual Landau prescription.
Equivalently, we have
\begin{equation}
k^{2}+\frac{1}{\lambda^{2}_{De}}=\frac{\omega^{2}_{\textrm{pi}}}{n_{0}}\int^{+\infty}_{-\infty}\frac{\partial f_{i0}/\partial v}{v-\omega/k}\,dv.
\end{equation}
where 
\begin{equation}
\lambda_{De}=\left(\frac{\epsilon_{0}T_{e}}{n_{0}e^{2}}\right)^{1/2},\quad\omega_{\textrm{pi}}=\left(\frac{n_{0}e^{2}}{m_{i}\epsilon_{0}}\right)^{1/2}.
\end{equation}
are the electron Debye length and ion plasma frequency, respectively.
The subscript `$0$' indicates an equilibrium quantity, and temperature
is expressed in energy units.

For a self-excited IICSI, we assume two counter-streaming population
each with velocity $\pm v_{b}/2$, so that our combined equilibrium
ion distribution function is given by 
\begin{equation}
f_{i0}=\frac{n_{0}}{2\sqrt{2\pi}v_{\textrm{Th}}}\left[\exp\left\{ -\frac{(v+v_{b}/2)^{2}}{2v^{2}_{\textrm{th}}}\right\} +\exp\left\{ -\frac{(v-v_{b}/2)^{2}}{2v^{2}_{\textrm{th}}}\right\} \right],
\end{equation}
where $v_{\textrm{th}}=\sqrt{T_{i}/m_{i}}$ is the ion thermal velocity.
This is a typical Landau damping (or growth) situation, which requires
a positive slope in the combined equilibrium function for a growth.
As our combined distribution function is symmetric around $v=0$,
for instability we need a minimum at $v=0$ for growth. So, the instability
condition reduces to 
\begin{equation}
\frac{d^{2}f_{i0}}{dv^{2}}>0.
\end{equation}
A simple calculation shows that this can be reduced to the condition
\citep{DasBora2025}
\begin{equation}
v_{b}>2v_{\textrm{th}}.
\end{equation}
However, there is a need for some clarifications. While deriving the
above condition, we have assumed the counter-streaming ion beam velocities
to be constant, which is \emph{not }true for a self-excited IICSI,
but $v_{b}\equiv v_{b}(x)$. Instead, we should have a condition 
\begin{equation}
v_{b,\textrm{max}}>2v_{\textrm{th}},\label{eq:ch11-ins-cond}
\end{equation}
where $v_{b,\textrm{max}}$ is the maximum velocity, an ion beam can
obtain due to debris potential. Let us now consider the debris potential
to be $\phi_{d}<0$, which causes an ion acceleration resulting the
velocity $v_{b}(x)$ with $x$ being the distance of the ion beam
position from the debris site. From energy conservation, we have 
\begin{equation}
\frac{1}{2}m_{i}\left[\frac{v_{b}(x)}{2}\right]^{2}=e|\phi(x)|,
\end{equation}
where $\phi(x)=\phi_{d}S(x)$, $S(x)$ being a \emph{shape factor},
which takes care of the local attenuation of the debris potential.
Thus 
\begin{equation}
v_{b}(x)=\left[\frac{8e|\phi_{d}|S(x)}{m_{i}}\right]^{1/2}.
\end{equation}
So, for a self-excited IICSI, we need the condition (\ref{eq:ch11-ins-cond})
at the minimum with 
\begin{equation}
v_{b,\textrm{max}}=\left(\frac{8e|\phi_{d}|}{m_{i}}\right)^{1/2}.
\end{equation}
From the PIC simulation \citep{DasBora2025}, it has already been
shown that the actual instability condition requires a much stricter
condition, which can be written as 
\begin{equation}
v_{b,\textrm{max}}>\eta v_{\textrm{Th}},\label{eq:vbmax}
\end{equation}
instead of relation (\ref{eq:ch11-ins-cond}) where $\eta>2$. To
understand this, we must examine the situation a bit more carefully.
In a self-excited instability, the growth rate \emph{must} be large
enough so that the instability can reach a nonlinear saturation regime
before the ions are convected away by their thermal motion, which
makes the factor $\eta$ larger than the linear limit of $2$. Following
the results of the PIC simulation \citep{DasBora2025}, we can safely
agree to have a value of $\eta\sim5-10$. So, the instability threshold
condition becomes 
\begin{equation}
e|\phi_{d}|>\frac{\eta^{2}}{8}T_{i}.
\end{equation}
So, the critical debris potential for excitation of the IICSI is 
\begin{equation}
|\phi_{d}|_{\textrm{crit}}=\frac{\eta^{2}}{8e}T_{i}.
\end{equation}
Assuming a spherical-shaped debris of radius $a$, the charge on the
debris is 
\begin{equation}
Q_{d}=4\pi\epsilon_{0}a\phi_{d},
\end{equation}
and the critical charge on the debris for a self-excited IICSI, we
have 
\begin{equation}
|Q_{d,\textrm{crit}}|=\pi\epsilon_{0}a\frac{\eta^{2}}{2e}T_{i},
\end{equation}
where $e$ is the numerical value of the electronic charge. Equivalently
we have 
\begin{equation}
Z_{d,\textrm{crit}}=\pi\epsilon_{0}a\frac{\eta^{2}}{2e^{2}}T_{i}
\end{equation}
as the critical charge number on the debris for a self excitation.

\subsubsection{The effect of ion mass}

Since $v_{b,\textrm{max}}=(8e|\phi_{d}|/m_{i})^{1/2}$ and $v_{\textrm{Th}}=(T_{i}/m_{i})^{1/2}$,
we have 
\begin{equation}
\frac{v_{b,\textrm{max}}}{v_{\textrm{Th}}}=\left(\frac{8e|\phi_{d}|}{T_{i}}\right)^{1/2}.
\end{equation}
Thus, at fixed $T_{i}$ and debris potential, the local threshold
$e|\phi_{d}|>\eta^{2}T_{i}/8$ is independent of ion mass. Ion mass
nevertheless changes the physical time and flow scales -- $\omega_{\textrm{pi}}\propto m^{-1/2}_{i}$,
and as mentioned earlier an $\textrm{O}^{+}$ plasma evolves four
times more slowly than an $\textrm{H}^{+}$ plasma when compared in
the same physical time unit. At a fixed physical orbital flow speed,
however, the sub and supersonic regimes become quite different for
$\textrm{H}^{+}$ and $\textrm{O}^{+}$. With $c_{s}\simeq(T_{e}/m_{i})^{1/2}$,
the Mach number scales as $M=u_{0}/c_{s}\propto m^{1/2}_{i}$. For
a representative equilibrium flow value of $u_{0}=7.5\,\textrm{km\,s}^{-1}$
and $T_{e}=1\,\textrm{eV}$, one approximately obtains $M=0.77$ for
$\textrm{H}^{+}$ and $M=3.06$ for $\textrm{O}^{+}$. Hence for an
$\textrm{O}^{+}$-dominated LEO plasma, the same orbital flow will
be in the high supersonic wake regime. However, as we use normalized
parameters in computation, a single Mach number can represent both
light and heavy ions. The plasma density also does not enter into
consideration as all densities are normalized by their equilibrium
values. As such, computational results with lighter ions are expected
to be same with heavier ions except that the latter evolves more slowly.

\subsubsection{Debris size}

The next \emph{most} important quantity is the size of the debris.
Consider a situation, when the size of the debris $a\gg\lambda_{De}$.
We argue that even with sufficient charge, this \emph{cannot} excite
an IICSI. As the debris potential (created at its boundary) is screened
within a very thin layer $\delta\sim\lambda_{De}\ll a$, all sides
of the debris develop its own local sheath. Although the ions will
be accelerated toward the sheath, they will not interact with other
ions streaming from the opposite direction and it reduces to a simple
independent sheath flow. For the ions to interact, we \emph{must }need
$a<\lambda_{De}$, so that the local sheath extends beyond the debris
size. With this condition, the conditions for a self-excited IICSI
are 
\begin{equation}
\begin{array}{rcl}
\phi_{d} & < & 0,\\
|\phi_{d}| & > & |\phi_{d}|_{\textrm{crit}},\\
a & < & \lambda_{De}.
\end{array}
\end{equation}
For $T_{i}\sim0.1\,\textrm{eV}$ and $\eta=5$, we have $Z_{d,\textrm{crit}}\sim10^{6}$.
Considering various regions of LEO plasma parameters, we find that
self-excited IICSI may be energetically plausible under some cold
topside/F-region conditions for millimetre- to centimetre-scale negatively
charged objects. Table~\ref{tab:ch11-LEO-parameter-range} summarizes
these order-of-magnitude estimates. In the table, 
\begin{equation}
f_{\mathrm{O}^{+}}=\frac{n_{\mathrm{O}^{+}}}{n_{\mathrm{O}^{+}}+n_{\mathrm{H}^{+}}}
\end{equation}
denotes the oxygen-ion number-density fraction. While $f_{\mathrm{O}^{+}}=1$
corresponds to a pure $\mathrm{O}^{+}$ plasma, $f_{\mathrm{O}^{+}}=0$
corresponds to a pure $\mathrm{H}^{+}$ plasma.

\begin{table}
\caption{LEO parameter range for self-excited IICSI \citep{Bilitza2017IRI,Bilitza2022IRI,Truhlik2005Composition,Truhlik2012ElectronTemperature,Pignalberi2021IonTemperature}.\vrr}
\label{tab:ch11-LEO-parameter-range}
\centering{}%
\begin{tabular}{|c|c|c|c|c|c|c|}
\hline 
LEO Region  & $T_{e}\,(\textrm{eV})$  & $f_{\mathrm{O}^{+}}$  & $T_{i}\,(\textrm{eV})$  & $\lambda_{D}$  & $|\phi_{d}|_{\textrm{crit}}\,(\textrm{V})$  & IICSI Status\vrr\tabularnewline
\hline 
\hline 
\begin{cellvarwidth}[t]
\centering
 $300-450\,\textrm{km}$\\
 ($F$-region peak) 
\end{cellvarwidth} & $0.08-0.45$  & $0.95$  & $0.06-0.15$  & $2-50\,\textrm{mm}$  & $0.19-0.47$\vrt  & \begin{cellvarwidth}[t]
\centering
 $\textrm{O}^{+}$ : viable

$\textrm{H}^{+}$ : unfavourable\vrd 
\end{cellvarwidth}\tabularnewline
\hline 
\begin{cellvarwidth}[t]
\centering
 $450-600\,\textrm{km}$\\
 (topside $F$-region) 
\end{cellvarwidth} & $0.10-0.60$  & $0.80$  & $0.07-0.20$  & $4-180\,\textrm{mm}$  & $0.22-0.63$  & \begin{cellvarwidth}[t]
\centering
 $\textrm{O}^{+}$ : favourable\vrt

$\textrm{H}^{+}$ : viable\vrd 
\end{cellvarwidth}\tabularnewline
\hline 
\begin{cellvarwidth}[t]
\centering
 $600-800\,\textrm{km}$\\
 (upper topside) 
\end{cellvarwidth} & $0.10-0.80$  & $0.50$  & $0.08-0.30$  & $7-300\,\textrm{mm}$  & $0.25-0.94$  & \begin{cellvarwidth}[t]
\centering
 $\textrm{O}^{+}$ : viable\vrt

$\textrm{H}^{+}$ : viable\vrd 
\end{cellvarwidth}\tabularnewline
\hline 
\begin{cellvarwidth}[t]
\centering
 $800-1000\,\textrm{km}$\\
 (upper LEO) 
\end{cellvarwidth} & $0.10-1.0$  & $0.30$  & $0.10-0.40$  & $10-750\,\textrm{mm}$  & $0.31-1.25$  & \begin{cellvarwidth}[t]
\centering
 $\textrm{O}^{+}$ : weakly viable\vrt

$\textrm{H}^{+}$ : favourable\vrd 
\end{cellvarwidth}\tabularnewline
\hline 
\end{tabular}
\end{table}

Table. \ref{tab:ch11-LEO-parameter-range} represents an order-of-magnitude
values rather than a specification of the parameters used in the simulations.
These parameters indicate that the cold-ion reference regime can be
energetically plausible for sufficiently negatively charged objects.
However, a calculation using detailed LEO composition and charging
conditions is beyond the scope of the present work.

\subsubsection{Debris charge}

The polarity of a millimeter-scale object in the topside ionosphere
depends on plasma collection, illumination, material properties, and
surface emission. Solar ultraviolet radiation can drive photoelectron
emission and cause positive charging, whereas the larger thermal electron
flux can result in negative charging. A simple current-balance estimate
is sufficient here to identify a plausible negative-charging regime.

Considering the primary mechanism for charging such debris, including
photoemission, we have the net current to the surface of a debris
through an OML-type current balance equation \citep{Whipple1981}
\begin{equation}
I_{e}(\phi_{d})+I_{i}(\phi_{d})+I_{\textrm{ph}}(\phi_{d})=0,
\end{equation}
where $I_{i}$, $I_{e}$, and $I_{\textrm{ph}}$ are the ion, electron,
and photoemission currents, respectively, and $\phi_{d}$ is the debris
potential. For a spherical grain (debris) of radius $a$, the zero-potential
electron current to the debris can be written as 
\begin{equation}
I_{e0}=en_{e}\pi a^{2}\sqrt{\frac{8T_{e}}{\pi m_{e}}},
\end{equation}
where $n_{e}$ is the electron density and $T_{e}$ is the electron
temperature expressed in energy units. The photoelectron current due
to solar UV photons can be written as 
\begin{equation}
I_{\textrm{ph}}\simeq J_{\textrm{ph}}A,
\end{equation}
where $J_{\textrm{ph}}$ is the photoelectron current density and
$A$ is the effective surface area of the grain. The ion current to
the debris can be expressed as 
\begin{equation}
I_{i}\sim en_{i}u_{0}\pi a^{2},
\end{equation}
where $u_{0}$ is the plasma flow past the debris. So, for a net negative
debris charge, we must have 
\begin{equation}
en_{e}\pi a^{2}\sqrt{\frac{8T_{e}}{\pi m_{e}}}>en_{i}u_{0}\pi a^{2}+J_{\textrm{ph}}A.
\end{equation}
Writing $J_{\textrm{ph}}A\equiv4\pi a^{2}J^{\textrm{eff}}_{\textrm{ph}}$,
the above condition becomes 
\begin{equation}
en_{e}\sqrt{\frac{8T_{e}}{\pi m_{e}}}>en_{i}u_{0}+J^{\textrm{eff}}_{\textrm{ph}}.
\end{equation}

Using a top-tier LEO ionospheric plasma, we can assume $n_{i}\sim n_{e}\sim10^{11}\,\textrm{m}^{-3}$,
$T_{e}\sim0.2\,\textrm{eV}$, $u_{0}\sim7.5\times10^{3}\,\textrm{m}\,\textrm{s}^{-1}$,
and $J^{\textrm{eff}}_{\textrm{ph}}\sim60\,\mu\textrm{A\,m}^{-2}$,
we can see that 
\begin{equation}
\left|\frac{I_{e}}{I_{i}+I_{\textrm{ph}}}\right|\sim30,
\end{equation}
indicating that negative charging is plausible for this illustrative
parameter set even when photoemission is included.

\subsubsection{Effect of collisions}

Although the plasma parameters used in the present calculation are
inspired by topside-ionospheric conditions, they are not intended
to reproduce a particular LEO plasma environment. Nevertheless, it
is necessary to examine the relevance of collisions with respect to
the specific physical regime considered here, namely the development
and nonlinear saturation of the debris-excited IICSI. In this section,
we primarily consider three types of collisions that are relevant
in the present context: ion--ion, electron--ion, and ion--neutral
collisions.

For a weakly coupled, singly ionized plasma, the Coulomb ion--ion
collision frequency may be estimated as \citep{Beresnyak2023} 
\begin{equation}
\nu_{ii}\simeq4.80\times10^{-8}\frac{n_{i}(\textrm{cm}^{-3})\ln\Lambda}{\sqrt{\mu_{i}}\,[T_{i}(\textrm{eV})]^{3/2}}\ \textrm{s}^{-1},
\end{equation}
where $\mu_{i}=m_{i}/m_{p}$ and $\ln\Lambda$ is the Coulomb logarithm.
The ion-plasma frequency can be written as 
\begin{equation}
\omega_{\textrm{pi}}\simeq1.32\times10^{3}\left[\frac{n_{i}(\textrm{cm}^{-3})}{\mu_{i}}\right]^{1/2}\ \textrm{s}^{-1},
\end{equation}
and hence 
\begin{equation}
\frac{\nu_{ii}}{\omega_{\textrm{pi}}}\simeq3.64\times10^{-11}\frac{\sqrt{n_{i}(\textrm{cm}^{-3})}\ln\Lambda}{[T_{i}(\textrm{eV})]^{3/2}}.
\end{equation}
As the ion-mass dependence cancels from this ratio, the relative importance
of Coulomb relaxation on the ion-plasma timescale is nearly the same
for $\textrm{H}^{+}$ and $\textrm{O}^{+}$ plasmas with equal densities,
charge states, and ion temperatures. For illustration, using $n_{i}=10^{5}\,\textrm{cm}^{-3}$,
$T_{i}=0.01\,\textrm{eV}$, and $\ln\Lambda\simeq10$, we have 
\begin{equation}
\frac{\nu_{ii}}{\omega_{\textrm{pi}}}\simeq1.2\times10^{-4}\ll1.
\end{equation}
Even if the density is increased by one order of magnitude, this ratio
remains below approximately $4\times10^{-4}$. Therefore, Coulomb
relaxation is much slower than the ion-plasma timescale and does not
modify the rapid collective ion response to leading order.

To obtain an order-of-magnitude estimate of $\gamma_{\mathrm{IICSI}}$,
we use the condition derived earlier, 
\begin{equation}
v_{b,\textrm{max}}>\eta v_{\textrm{Th}},\qquad\eta\simeq5-10,\label{eq:vbmax-collisions}
\end{equation}
which implies that the two ion populations are well separated in velocity
space. Since their individual drift velocities are approximately $\pm V$,
where $V=v_{b}/2$, we have 
\begin{equation}
\frac{V}{v_{\textrm{Th}}}>\frac{\eta}{2}\simeq2.5-5.
\end{equation}
The cold-beam approximation can therefore be used to estimate the
characteristic scale of the instability growth rate. For two symmetric
cold-ion populations, each having density $n_{0}/2$ and equilibrium
velocities $\pm V$, with Boltzmann electrons, the electrostatic fluid
dispersion relation is 
\begin{equation}
1+\frac{1}{k^{2}\lambda^{2}_{De}}-\frac{\omega^{2}_{\textrm{pi}}}{2}\left[\frac{1}{(\omega-kV)^{2}}+\frac{1}{(\omega+kV)^{2}}\right]=0.\label{eq:cold-two-beam-dispersion}
\end{equation}
Defining 
\begin{equation}
\Omega^{2}_{k}=\frac{\omega^{2}_{\textrm{pi}}}{1+1/(k^{2}\lambda^{2}_{De})}=\omega^{2}_{\textrm{pi}}\frac{k^{2}\lambda^{2}_{De}}{1+k^{2}\lambda^{2}_{De}},
\end{equation}
together with 
\begin{equation}
y=\frac{\omega^{2}}{\Omega^{2}_{k}},\qquad a=\frac{k^{2}V^{2}}{\Omega^{2}_{k}},
\end{equation}
Eq.~(\ref{eq:cold-two-beam-dispersion}) reduces to 
\begin{equation}
(y-a)^{2}-(y+a)=0.
\end{equation}
The unstable branch has $\omega=i\gamma$, and hence $y=-\gamma^{2}/\Omega^{2}_{k}$.
Therefore, 
\begin{equation}
\frac{\gamma^{2}}{\Omega^{2}_{k}}=\frac{\sqrt{1+8a}-1-2a}{2},\qquad0<a<1.
\end{equation}
It can be easily verified that the maximum growth occurs at $a=3/8$,
giving 
\begin{equation}
\gamma_{\max}=\frac{\Omega_{k}}{2\sqrt{2}}\simeq0.35\,\omega_{\textrm{pi}}\frac{k\lambda_{De}}{\sqrt{1+k^{2}\lambda^{2}_{De}}}.\label{eq:iicsi-growth-estimate}
\end{equation}
For the Debye-scale and several-Debye-length structures relevant to
the present precursor and wake, a representative range may be taken
as $k\lambda_{De}\sim0.3-1$. Equation~(\ref{eq:iicsi-growth-estimate})
then gives 
\begin{equation}
\gamma_{\max}\sim(0.1-0.25)\omega_{\textrm{pi}}.
\end{equation}
Combining the earlier result with this relation, we find 
\begin{equation}
\frac{\nu_{ii}}{\gamma_{\max}}\sim5\times10^{-4}-10^{-3}\ll1.
\end{equation}
Thus, ion-ion relaxation is much slower than the characteristic cold-beam
growth and early nonlinear development of the IICSI in the working
regime considered here. However, $\gamma_{\max}$ is an order-of-magnitude,
cold-beam estimate rather than the actual growth rate of the finite-temperature,
spatially inhomogeneous ion populations in the PIC simulation.

For electron-ion collisions, the frequency may similarly be estimated
from 
\begin{equation}
\nu_{ei}\simeq2.91\times10^{-6}\frac{n_{e}(\textrm{cm}^{-3})\ln\Lambda}{[T_{e}(\textrm{eV})]^{3/2}}\ \textrm{s}^{-1}.
\end{equation}
For the electron-heated regime used in the present calculation, $\nu_{ei}$
is small compared with both $\omega_{\textrm{pe}}$ and the characteristic
IICSI growth scale estimated above. Electron-ion collisions therefore
do not alter the rapid electron shielding response to leading order.

Ion-neutral collisions require separate consideration because their
frequency depends on the neutral density and the momentum-transfer
cross-section,
\begin{equation}
\nu_{in}=n_{n}\left\langle \sigma_{in}v_{\mathrm{rel}}\right\rangle .
\end{equation}
The neutral density decreases strongly in the topside ionosphere,
and ion--neutral collisions are generally much weaker there than
in the lower ionosphere \citep{Ieda2021}. However, because the present
calculation is not associated with a specified altitude, local time,
or neutral-atmospheric state, a unique value of $\nu_{in}$ cannot
be assigned. The present results therefore apply to the weakly collisional
topside regime satisfying 
\begin{equation}
\frac{\nu_{in}}{\gamma_{\mathrm{IICSI}}}\ll1.
\end{equation}
Cases in which ion--neutral damping becomes comparable with the instability
growth rate lie outside the working regime of this study.

Based on the above discussion, the working ordering adopted in this
proof-of-principle calculation is 
\begin{equation}
\nu_{ii},\,\nu_{ei},\,\nu_{in}\ll\gamma_{\mathrm{IICSI}}\lesssim\omega_{\textrm{pi}}\ll\omega_{\textrm{pe}}.
\end{equation}
The estimates of the collision frequencies and the characteristic
cold-beam growth scale show that this ordering is physically plausible
for a weakly collisional topside-ionospheric regime. Under this ordering,
the formation and nonlinear development of the IICSI are governed
primarily by collective wave-particle dynamics and collisions may
influence the plasma \emph{only }over a longer transport and relaxation
time, but they do not enter the leading-order mechanism examined in
the present work.

\begin{table}
\caption{PINN parameters (left) and the discovered effective parameters at
$t=10\omega^{-1}_{\text{pi}}$ and $t=20\omega^{-1}_{\text{pi}}$
(right).\vrd}\label{tab:Parameters-value-for}

\centering{}%
\begin{tabular}[t]{|c|c|}
\hline 
PINN parameters & Value\vrr\tabularnewline
\hline 
\hline 
Inputs & $2$ ($x$ and $t$)\vrr\tabularnewline
\hline 
Output & $1$ ($n_{i}$)\vrr\tabularnewline
\hline 
Hidden layers & $8$\vrr\tabularnewline
\hline 
Neurons/layer & $64$\vrr\tabularnewline
\hline 
Activation function & $\tanh$\vrr\tabularnewline
\hline 
Optimizer & Adam\vrr\tabularnewline
\hline 
\end{tabular}\qquad{}%
\begin{tabular}[t]{|c|c|c|}
\hline 
Parameters & $t=10\omega^{-1}_{\text{pi}}$ & $t=20\omega^{-1}_{\text{pi}}$\vrr\tabularnewline
\hline 
\hline 
$A$ & $0.59$ & $0.46$\vrr\tabularnewline
\hline 
$B$ & $0.03$ & $0.01$\vrr\tabularnewline
\hline 
$C$ & $0.13$ & $0.02$\vrr\tabularnewline
\hline 
\end{tabular}
\end{table}

\section{Failure of a reduced fKdV description}\label{sec:Failure-of-a}

Forced-Korteweg-de Vries (fKdV) models can describe weakly nonlinear,
dispersive structures driven by a moving charged object and have been
used to examine possible signatures of sub-centimeter orbital debris
\citep{Truitt2020}. Such a reduced description is useful only while
the kinetic response remains in the early development stage and increasingly
becomes less reliable as the response attains a full-blown nonlinear
statistical equilibrium. We therefore begin by testing its range of
validity against PIC data.

We assume that the nonlinear response during a plasma-debris interaction
through IICSI can, in principle, be modeled by a generic fKdV equation
in ion density $n_{i}$, 
\begin{equation}
\frac{\partial n_{i}}{\partial t}+An_{i}\frac{\partial n_{i}}{\partial x}+B\frac{\partial^{3}n_{i}}{\partial x^{3}}=C\frac{\partial}{\partial x}\rho_{\textrm{ext}},\label{eq:fkdv}
\end{equation}
where 
\begin{equation}
\rho_{\textrm{ext}}=S_{0}\exp\left[-\frac{(x-v_{d}t)^{2}}{2\delta^{2}}\right]
\end{equation}
is the charge density of the external debris. The external charge
density is a Gaussian-shaped function where $S_{0}$ denotes the strength
of the debris charge, $v_{d}$ is the constant velocity of the debris
with respect to the ambient plasma, and $\delta$ determines the width
of the Gaussian. Here, $S_{0}$, $v_{d}$, and $\delta$ are normalized
by the equilibrium plasma density $n_{0}$, ion-acoustic speed $c_{s}$,
and electron Debye length $\lambda_{De}$ respectively. For the present
analysis, we use $S_{0}=-0.5$, $v_{d}=0.5$, and $\delta=0.6$ in
these normalized units. The effective parameters $A,B,C$ are inferred
using a physics-informed neural network (PINN) \citep{Raissi2019PINN,Karniadakis2021PIML},
while the parameters $S_{0},v_{d},\delta$ are fed into the PINN model.
We use a one-dimensional periodic PIC framework as our experimental
testbed, data from which will be used to discover the unknown parameters
of the fKdV equation using PINN.

Our PINN model is trained using the ion density $(n_{i})$ PIC data.
The network architecture is summarized in Table~\ref{tab:Parameters-value-for}
(left). Its inputs are the spatial coordinate $x$ and time $t$,
and its output $\widehat{n}_{i}(x,t)$ approximates the PIC ion density
$n_{i}(x,t)$. The approximate $\widehat{n}_{i}(x,t)$ is then differentiated
and derivatives are plugged into the fKdV equation with the residual
\begin{equation}
r_{\textrm{PINN}}=\frac{\partial\widehat{n}_{i}}{\partial t}+A\widehat{n}_{i}\frac{\partial\widehat{n}_{i}}{\partial x}+B\frac{\partial^{3}\widehat{n}_{i}}{\partial x^{3}}-C\frac{\partial}{\partial x}\left\{ S_{0}\exp\left[-\frac{(x-v_{d}t)^{2}}{2\delta^{2}}\right]\right\} .
\end{equation}

The neural network then computes the following loss functions 
\begin{eqnarray}
L_{\textrm{PDE}} & = & \frac{1}{N_{c}}\sum^{N_{c}}_{i=1}\left|r_{\textrm{PINN}}(x_{i},t_{i})\right|^{2},\\
L_{\textrm{data}} & = & \frac{1}{N_{d}}\sum^{N_{d}}_{i=1}\left|\widehat{n}_{i}(x_{i},t_{i})-n_{i}(x_{i},t_{i})\right|^{2},
\end{eqnarray}
where $N_{c}$ and $N_{d}$ are the numbers of collocation and data
points, respectively. Training minimizes $L=L_{\textrm{PDE}}+L_{\textrm{data}}$
using Adam, followed by limited-memory Broyden--Fletcher--Goldfarb--Shanno
(L-BFGS) refinement \citep{LiuNocedal1989LBFGS}.

\begin{figure}[t]
\begin{centering}
\includegraphics[width=0.495\textwidth]{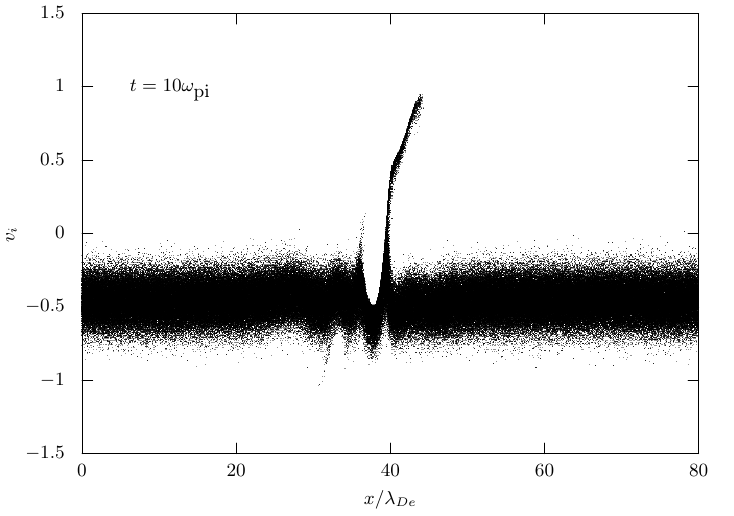}\hfill{}\includegraphics[width=0.495\textwidth]{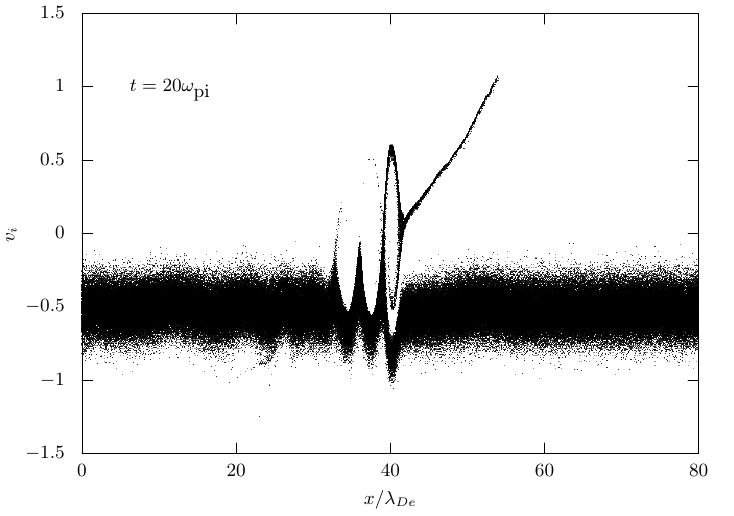}
\par\end{centering}
\caption{Ion phase space at $\omega_{\textrm{pi}}t=10$ (left) and $20$ (right).
The left panel shows the developing counter-streaming response, whereas
the right panel shows the fully developed nonlinear state.}
\label{fig:Ion-phase-space-at}
\end{figure}

\begin{figure}[t]
\begin{centering}
\includegraphics[width=0.495\textwidth]{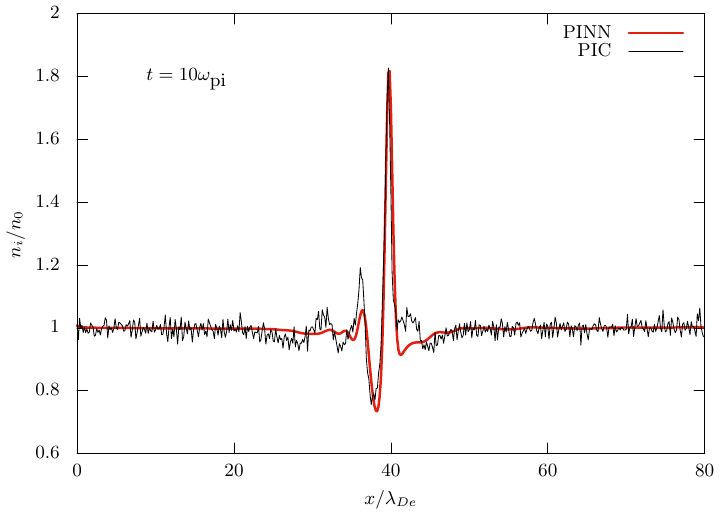}\hfill{}\includegraphics[width=0.495\textwidth]{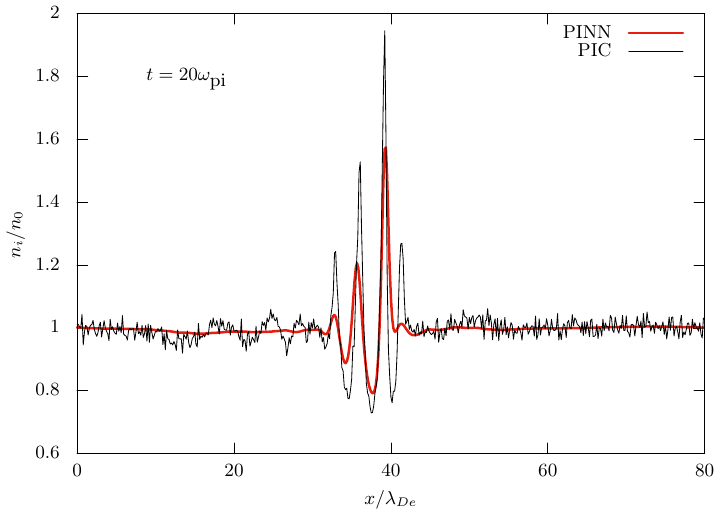} 
\par\end{centering}
\caption{PINN-constrained fKdV reconstruction of the PIC ion density at $t=10\omega^{-1}_{\textrm{pi}}$
(left) and $20\omega^{-1}_{\textrm{pi}}$ (right). The deterioration
at the later time marks the limitation of the reduced fKdV description.}
\label{fig:PINN-prediction-of} 
\end{figure}

For this preliminary test, the external charge is fixed at the value
specified above, for which the counter-streaming instability develops.
The phase-space snapshots in Fig.~\ref{fig:Ion-phase-space-at} show
the developing response at $t=10\omega^{-1}_{\textrm{pi}}$ and its
nonlinear state at $t=20\omega^{-1}_{\textrm{pi}}$. During the earlier
stage, the fitted fKdV model reproduces the PIC density reasonably
well {[}Fig.~\ref{fig:PINN-prediction-of}(a){]}. The discovered
values of the effective parameters $A$, $B$, and $C$ at this stage
are listed in Table~\ref{tab:Parameters-value-for}. Once phase-space
vortices and a statistically saturated kinetic response have developed,
the reduced model no longer reproduces the PIC density {[}Fig.~\ref{fig:PINN-prediction-of}(b){]}.
The comparison is not intended as a general assessment of PINNs, rather
it shows that a single weakly nonlinear fKdV closure is insufficient
for the regime from which the localization signatures are extracted.
This motivates the open-boundary kinetic treatment used in the next
section.

\section{Open-boundary PIC configuration}\label{sec:Open-boundary-PIC-configuration}

We next describe the open-boundary implementation used to simulate
a continuous one-dimensional plasma flow. It is implemented within
the \emph{hybrid}-PIC--MCC (\emph{h}-PIC--MCC) code \citep{ChangmaiBora2019,ChangmaiBora2020}.
The PIC and Monte Carlo collision methodology follows the standard
particle-simulation framework \citep{Birdsall1991,Verboncoeur2005}.

\subsection{Normalization}

We use the following electron-plasma normalization scheme 
\begin{align}
t & =\omega_{\textrm{pe}}t_{\mathrm{phys}}, & x & =x_{\mathrm{phys}}/\lambda_{De}, & v & =v_{\mathrm{phys}}/v_{\textrm{th}e0},\\
n & =n_{\mathrm{phys}}/n_{0}, & \phi & =e\phi_{\mathrm{phys}}/T_{e0}, & E & =E_{\mathrm{phys}}/(\phi_{0}/\lambda_{De}),
\end{align}
where the subscript `phys' denotes physical parameters in SI. As seen,
the time is normalized by the inverse of the electron plasma frequency,
length by electron Debye length, density by the equilibrium density
$n_{0}$, and electrostatic potential energy by the equilibrium thermal
energy where temperature is written in the energy unit. Specifically,
we have the following standard definitions for the plasma parameters.
\begin{equation}
\omega_{\textrm{pe}}=\sqrt{\frac{n_{0}e^{2}}{m_{e}\epsilon_{0}}},\quad\lambda_{De}=\frac{v_{\textrm{th}e0}}{\omega_{\textrm{pe}}},\quad v_{\textrm{th}e0}=\sqrt{\frac{T_{e0}}{m_{e}}},\quad\phi_{0}=\frac{T_{e0}}{e},\quad T_{s}=\frac{T_{s,\textrm{phys}}}{T_{e0}},
\end{equation}
where $T_{s}$ is the normalized temperature of species $s$. Mass
and charge of a particle are normalized by the electronic mass $m_{e}$
and value of the electronic charge $e$. Naturally, in these normalized
units, we have $m_{e}=1$, $q_{e}=-1$, $q_{i}=+1$, and $\epsilon_{0}=1$.
The one-dimensional Poisson equation is given by 
\begin{equation}
\frac{\partial^{2}\phi}{\partial x^{2}}=n_{e}-n_{i}-\rho_{d},
\end{equation}
where $\rho_{d}$ is the signed normalized charge density of the external
debris. For any species $s$, the computational particle has the weight
$w_{s}$ 
\begin{equation}
w_{s}=\frac{n_{s}L}{N_{s}},
\end{equation}
where $N_{s}$ is the number of computational particles initially
used for that species and $L$ is the simulation domain length.

\subsection{Plasma flow}

The code is designed to simulate a continuous average equilibrium
plasma flow $u_{0}$ across the simulation domain expressed through
a normalized flow speed $M$, the Mach number 
\begin{equation}
u_{0}=\frac{M}{\sqrt{m_{i}/m_{e}}},
\end{equation}
which is positive along the $x$-axis i.e.\ the flow is from left
to the right of the simulation domain. The ion-acoustic speed used
for this conversion is $1/\sqrt{m_{i}/m_{e}}$.

\subsection{Open-boundary and end points}

\subsubsection{Grid configuration}

The common numerical configuration of the open-boundary PIC simulation
is listed in Table. \ref{tab:pic-base-parameters} (Appendix A). Quantities varied between individual
runs, including the flow Mach number, debris position, and run duration,
are specified with the corresponding cases and are not part of this
baseline parameter set.

The simulation domain interval $[0,L]$ is divided into a grid of
equal sized cells of width $\Delta x$, 
\begin{equation}
x_{j}=(j-1)\Delta x,\qquad j=1,\ldots,N_{x},\qquad\Delta x=\frac{L}{N_{x}-1},
\end{equation}
where $x_{1}=0$ and $x_{N_{x}}=L$. Particle density is calculated
at the grid points through the standard cloud-in-cell interpolation.
After interpolation, the first and last nodal densities are multiplied
by two because their nodal volumes have width $\Delta x/2$, rather
than $\Delta x$. Field interpolation back to particles uses the same
linear shape function.

\subsubsection{Infinite reservoir and sink}

We model the open-boundary with an infinite particle reservoir and
a sink for an equilibrium plasma flow. In the following discussion,
we shall assume a positive plasma flow, from left to right.

At time $t=0$, each species is initialized throughout the domain
with randomized positions 
\begin{equation}
x_{p}=L\left(\frac{p-1+\xi_{p}}{N_{s}}\right),\qquad\xi_{p}\sim{\cal U}(0,1),
\end{equation}
Initial velocities are sampled with a drifting Maxwellian, 
\begin{equation}
v_{p}=u_{0}+v_{\textrm{th},s}Z_{p},\qquad Z_{p}\sim\mathcal{N}(0,1),
\end{equation}
with 
\begin{equation}
v_{\textrm{th},e}=\sqrt{T_{e}},\qquad v_{\textrm{th},i}=\sqrt{\frac{T_{i}}{m_{i}/m_{e}}}.
\end{equation}
In the above expressions, ${\cal U}(0,1)$ denotes the continuous
uniform probability distribution between $0$ and $1$ and $\mathcal{N}(0,1)$
is a random number drawn from the standard normal probability distribution.

After a drift, any particle for which $x<0$ or $x>L$, is removed
from the simulation. This `drift' includes the particle's complete
velocity, including bulk, thermal, and field-accelerated contributions.
Particles exactly at $0$ or $L$ remain valid. The normalized charge
for a species $s$ is accumulated accordingly 
\begin{equation}
Q_{\mathrm{escaped},s}=N_{\mathrm{removed},s}q_{s}w_{s}.
\end{equation}
However the incoming reservoir injection is calculated separately
for electrons and ions at each active boundary, assuming a drifting
Maxwellian 
\begin{equation}
f_{s}(v)=\frac{n_{s}}{\sqrt{2\pi}v_{\textrm{th},s}}\exp\left[-\frac{(v-u_{0})^{2}}{2v^{2}_{\textrm{th},s}}\right].
\end{equation}
Then, the incoming number fluxes are given by 
\begin{align}
\Gamma_{s,\textrm{left}} & =\int^{\infty}_{0}vf_{s}(v)\,dv,\label{eq:fluxl}\\
\Gamma_{s,\textrm{right}} & =\int^{0}_{-\infty}(-v)f_{s}(v)\,dv.\label{eq:fluxr}
\end{align}
Apparently these integrals can be broken into two parts -- the thermal
part and the equilibrium drift part. For example, introducing the
variables $z=(v-u_{0})/v_{\textrm{th},s}$ and $a_{s}=u_{0}/v_{\textrm{th},s}$,
$\Gamma_{s,\textrm{left}}$ can be written as 
\begin{equation}
\Gamma_{s,\textrm{left}}=n_{s}u_{0}\left(\frac{1}{\sqrt{2\pi}}\int^{\infty}_{-a_{s}}e^{-z^{2}/2}dz\right)+n_{s}v_{\textrm{th},s}\left(\frac{1}{\sqrt{2\pi}}\int^{\infty}_{-a_{s}}ze^{-z^{2}/2}dz\right).
\end{equation}
Writing the first term in the bracket as $\varrho(a_{s})$ and the
second as $\psi(a_{s})$, Eqs. (\ref{eq:fluxl}) and (\ref{eq:fluxr})
can be written as 
\begin{align}
\Gamma_{s,\textrm{left}} & =\int^{\infty}_{0}vf_{s}(v)\,dv=n_{s}\left[v_{\textrm{th},s}\psi(a_{s})+u_{0}\varrho(a_{s})\right],\\
\Gamma_{s,\textrm{right}} & =\int^{0}_{-\infty}(-v)f_{s}(v)\,dv=n_{s}\left[v_{\textrm{th},s}\psi(a_{s})-u_{0}\varrho(-a_{s})\right].
\end{align}
The quantities $\psi$ and $\varrho$ are the standard-normal probability
density and cumulative distribution functions of the particles. The
left reservoir supplies particles with $v>0$ and the right reservoir
supplies particles with $v<0$. So, these fluxes give the number of
particles entering the system by using a drifting Maxwellian which
calculates the bulk flow and small individual deviations in velocities
of the particles due to random thermal motion. We \emph{must} note
that the infinite sink is \emph{not} modeled through these fluxes,
as mentioned earlier, the infinite sink just removes the particles
whenever they go out of the domain $[0,L]$ either due to random drift
or due to bulk flow. This way, this open-boundary configuration mimics
the equivalence of the simulation domain to a realistic observation
window in a flowing plasma.

A specific case arises for very strong equilibrium flow, when 
\begin{equation}
a_{s}=\frac{u_{0}}{v_{\textrm{th},s}}\gg1,
\end{equation}
i.e.\ the bulk flow is then much stronger than the thermal speed.
In this case, almost no particles at the right reservoir possess the
negative velocity needed to enter the domain. Accordingly, the right-boundary
flux will be extremely small 
\begin{equation}
\Gamma_{s,\textrm{right}}\ll1\Rightarrow v_{\textrm{th},s}\psi(a_{s})\approx u_{0}\varrho(-a_{s}),\label{eq:mills-flux}
\end{equation}
and the magnitudes of each individual terms are also very small, which
is extremely erroneous numerically and creates a catastrophic cancellation
in floating-point arithmetic. To circumvent this, small difference
in these formulas is evaluated with a Mills-ratio asymptotic expansion
rather than by subtracting nearly equal floating-point numbers (see
Appendix~B).

\subsubsection{Accumulation and buffer zones}

An open-boundary simulation is prone to artificial numerical effects
caused by particles entering and leaving at the boundaries. The buffer
zones are therefore used to prevent these boundary effects from contaminating
the physical interior of the domain.

For a species $s$, boundary $b$ (right or left), and one time step,
the expected number of injected computational particles is given by
\begin{equation}
\eta_{s,b}=\frac{\Gamma_{s,b}\Delta t}{w_{s}}.
\end{equation}
Note that $\eta_{s,b}$ is \emph{not} an integer.\emph{ }The \emph{h}-PIC-MCC
code thus maintains four fractional accumulators -- electron-left,
electron-right, ion-left, and ion-right. At every step, $\eta_{s,b}$
is added to the corresponding accumulator, its integer part is injected,
and its fractional remainder is retained. This avoids systematic rounding
error in the long-time reservoir flux. When particles are added, their
velocities are also needed to be assigned. This is done through flux-weighted
probability distributions 
\begin{eqnarray}
p_{s,\textrm{left}}(v) & = & \frac{vf_{s}(v)}{\Gamma_{s,\textrm{left}}},\quad v>0,\\
p_{s,\textrm{right}}(v) & = & \frac{(-v)f_{s}(v)}{\Gamma_{s,\textrm{right}}},\quad v<0,
\end{eqnarray}
As the sampler uses rejection sampling, for particles moving opposite
to a strong equilibrium flow, we adopt a Rayleigh-distributed speed
proposal, as explained below.

Assume that there is a particle moving from right to left (entering
the domain) against a very strong equilibrium flow from left to the
right. In our case, this is the case when 
\begin{equation}
u_{0}\gg v_{\textrm{th},s},\quad u>0,\quad v<0.
\end{equation}
Due to strong exponential dependence of the drifting Maxwellian, random
sampling of the full Maxwellian for velocity assignment is extremely
inefficient. Instead, we write the incoming speed as 
\begin{equation}
w=|v|>0.
\end{equation}
For a right-boundary injection, the flux-weighted distribution is
given by 
\begin{equation}
p(w)\propto w\,\exp\left[-\frac{(w+u_{0})^{2}}{2v^{2}_{\textrm{th},s}}\right]\equiv w\,\exp\left(-\frac{w^{2}}{2v^{2}_{\textrm{th},s}}\right)\exp\left(-\frac{u_{0}w}{v^{2}_{\textrm{th},s}}\right).
\end{equation}
In the above expression, the first exponential part (with $w$) is
the Rayleigh proposal \citep{Devroye1986RandomVariates} and the second
one is the acceptance factor. The code therefore draws a positive
speed $w$ from a Rayleigh distribution 
\begin{equation}
w=v_{\textrm{th},s}\sqrt{-2\,\ln U_{1}},\quad U_{1}\sim{\cal U}(0,1)
\end{equation}
and accepts it with a probability 
\begin{equation}
{\cal P}_{\textrm{accept}}=\exp\left(\frac{-|u_{0}|w}{v^{2}_{\textrm{th},s}}\right).
\end{equation}
$U_{1}$ is a random uniform deviate. After acceptance, it re-assigns
the legitimate sign i.e.\ $v=-w$. Besides, an injected particle
is placed inside the distance that it could have traveled during the
current time step 
\begin{align}
x_{\mathrm{new}} & =\xi v\Delta t, &  & \text{left boundary, }v>0,\\
x_{\mathrm{new}} & =L-\xi|v|\Delta t, &  & \text{right boundary, }v<0,
\end{align}
where $\xi\sim{\cal U}(0,1)$.

The \emph{h}-PIC-MCC code also has provisions for boundary buffers
(also known as `sponge'), which occupies 
\[
0\leq x<L_{\textrm{buffer}},\qquad L-L_{\textrm{buffer}}<x\leq L,
\]
where $L_{\textrm{buffer}}$ is the buffer length. This defines the
depth, measured inward from the inner edge of a buffer zone by 
\begin{equation}
d(x)=\begin{cases}
L_{\textrm{buffer}}-x, & x<L_{b},\\
x-(L-L_{\textrm{buffer}}), & x>L-L_{b},\\
0, & \text{otherwise}.
\end{cases}
\end{equation}
We introduce a quantity called the local relaxation rate $\alpha(x)$,
which measures how rapidly the buffer drives particle velocities toward
the prescribed reservoir Maxwellian at position $x$, 
\begin{equation}
\alpha(x)=\alpha_{\max}\left(\frac{d(x)}{L_{\textrm{buffer}}}\right)^{\kappa},
\end{equation}
where $\kappa$ is an exponent which determines how the relaxation
strength varies across the buffer region. So, 
\begin{equation}
\begin{array}{rclll}
\kappa & = & 1 & : & \mbox{linear increase across the buffer},\\
 & = & 2 & : & \mbox{quadratic, weak at the interior, strong near boundary},\\
 & > & 2 & : & \mbox{weak but rises sharply near boundary},\\
 & \in & (0,1) & : & \mbox{becomes strong immediately inside}.
\end{array}
\end{equation}
In our simulation, we use $\kappa=2$, $\alpha_{\textrm{max}}=0.2$,
and $L_{\textrm{buffer}}=20$. The buffer is applied to electrons
and ions after a particle drift, debris collection, removal of escaped
particles, and reservoir injection. For each particle currently inside
either buffer zone, the code calculates a reset probability 
\begin{equation}
{\cal P}_{\mathrm{reset}}(x)=1-\exp[-\alpha(x)\Delta t].
\end{equation}
If a Bernoulli trial with this reset probability succeeds, the velocity
of the particle is replaced by 
\begin{equation}
v\leftarrow u_{0}+v_{\textrm{th},s}Z,\quad Z\sim{\cal N}(0,1).
\end{equation}
The buffer zone therefore acts only within 20 normalized distance
units of the two physical boundaries. It affects particle velocities
only. It is not applied to the interior particles, density, potential,
electric field, particle position, charge, or particle weight.

The code is designed to have an arbitrary domain length with proportionally
scaled up (or down) computational parameters so as to keep the plasma
parameters unchanged. This way, we can practically increase the spatial
observation window (domain length $L$) so that enough observable
regions are included for wake and precursor diagnostics (see Section
\ref{sec:Precursor-and-wake}) before the particles encounter the
open-boundary reservoir and sink.

\subsubsection{Open-boundary Poisson solver}

As our boundaries are open, we apply Dirichlet conditions to the electrostatic
potential $\phi$, 
\begin{equation}
\phi(0)=\phi_{L},\qquad\phi(L)=\phi_{R}.
\end{equation}
For interior nodes, the finite-difference equation is 
\begin{equation}
-\phi_{j-1}+2\phi_{j}-\phi_{j+1}=\Delta x^{2}\rho_{j},\qquad j=2,\ldots,N_{x}-1.
\end{equation}
The known endpoint values are moved to the first and last entries
of the right-hand side. The resulting tridiagonal system is solved
directly. In contrast to a periodic Poisson solver, no spatial mean
of $\rho$ is subtracted.

The electric field is calculated with centered differences in the
interior and one-sided differences at the endpoints: 
\begin{align}
E_{1} & =-\frac{\phi_{2}-\phi_{1}}{\Delta x},\\
E_{j} & =-\frac{\phi_{j+1}-\phi_{j-1}}{2\Delta x},\qquad2\leq j\leq N_{x}-1,\\
E_{N_{x}} & =-\frac{\phi_{N_{x}}-\phi_{N_{x}-1}}{\Delta x}.
\end{align}
The boundary is therefore open to particles but electrostatically
closed by the specified potentials. In the present configuration,
no external voltage is imposed across the simulation domain, and therefore,
the boundary potentials are set to zero 
\begin{equation}
\phi_{L}=\phi_{R}=0.
\end{equation}

\section{Debris interaction within the open domain}\label{sec:Debris-interaction-within}

The debris center is user-defined through a fraction $f_{d}$ of the
domain length $L$, 
\begin{equation}
x_{d}=f_{d}L.
\end{equation}
The spatial charge collection window by the external debris is $[x_{d}-R_{d},x_{d}+R_{d}]$,
which must lie inside the domain, where $R_{d}$ is the `radius' surrounding
the center of the debris over which collection takes place. The debris
charge distribution is assumed to be Gaussian and its charge is represented
by markers distributed over that interval. Their unnormalized Gaussian
weights are 
\begin{equation}
g_{k}=\exp\left[-\frac{1}{2}\left(\frac{x_{k}-x_{d}}{\sigma_{d}}\right)^{2}\right],\qquad\sigma_{d}=f_{\sigma}R_{d},
\end{equation}
and their normalized fractions are $g_{k}/\sum_{j}g_{j}$, so that
the marker charges sum exactly to the instantaneous total debris charge
accumulated up to that time. Marker charge is deposited by cloud-in-cell
interpolation into $\rho_{d}$.

The charge collection is tested on the complete particle trajectory
from its pre-drift to its post-drift position. Thus, a fast particle
cannot pass across the debris interval without being tested. For a
species-specific kinetic energy $K=mv^{2}/2$, the capture probability
is 
\begin{equation}
{\cal P}_{\mathrm{cap}}(K)=\begin{cases}
0, & K\leq E_{\mathrm{th}},\\
{\cal P}_{\max}\left[1-\exp\left(-\dfrac{K-E_{\mathrm{th}}}{E_{\mathrm{scale}}}\right)\right], & K>E_{\mathrm{th}},
\end{cases}
\end{equation}
where $E_{\textrm{th}}$ is threshold energy and $E_{\textrm{scale}}$
controls how quickly the capture probability rises above the threshold.
A particle with $K\leq E_{\textrm{th}}$ is \emph{not }captured. In
the present simulation 
\begin{equation}
\begin{array}{rcl}
{\cal P}_{\textrm{max},e} & = & 0.01,\\
{\cal P}_{\textrm{max},i} & = & 0.005,
\end{array}\quad\begin{array}{rcl}
E_{\textrm{scale},e} & = & 0.2,\\
E_{\textrm{scale},i} & = & 0.05,
\end{array}
\end{equation}
and $E_{\textrm{th}}=0$ for both electrons and ions.

The capture model above is a phenomenological numerical prescription
rather than a first-principles, material-specific collection law.
It is nevertheless motivated by the physical expectation that the
probability of surface collection increases with the kinetic energy
of an incident particle above an effective threshold and approaches
a finite limiting value. The separate electron and ion parameters
represent their different collection efficiencies. Their values can
be varied to control the charging rate and the resulting statistically
stationary charge; they should therefore be regarded as model parameters,
not universal properties of orbital debris. In the present \emph{proof-of-principle}
calculation, this prescription provides a controlled means of generating
self-consistent charge evolution while retaining the essential feedback
between particle collection, debris charge, and the surrounding plasma
response.

\subsection{Debris-excited IICSI and charge fluctuation}

\begin{figure}[t]
\begin{centering}
\includegraphics[width=0.495\textwidth]{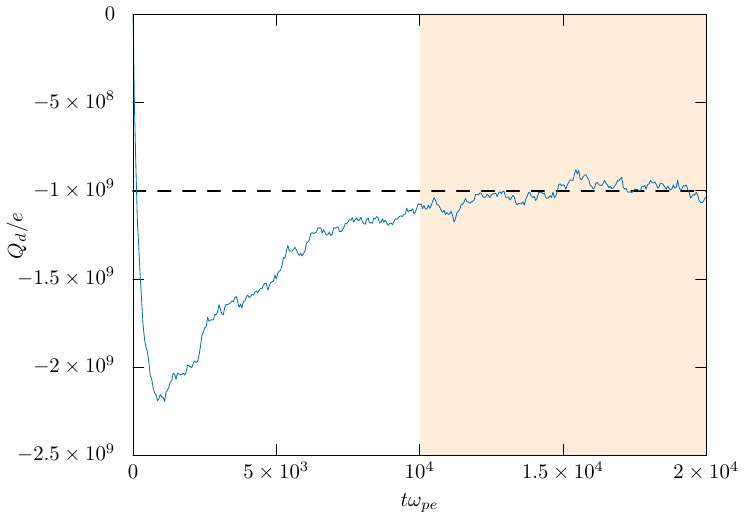}\hfill{}\includegraphics[width=0.495\textwidth]{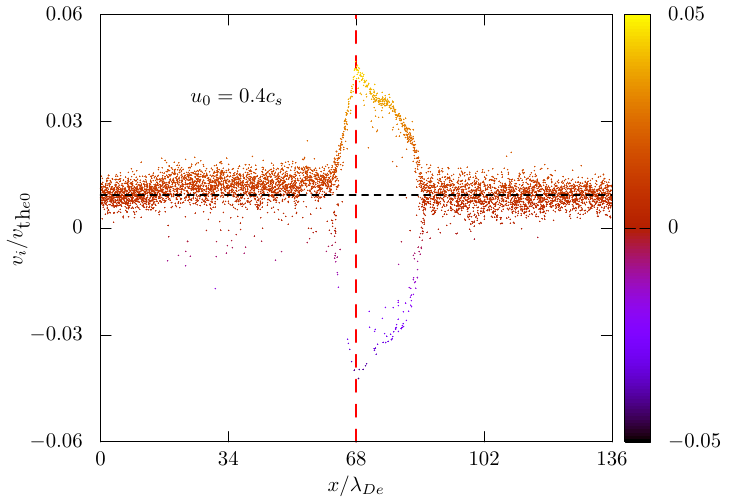} 
\par\end{centering}
\caption{Left:~Evolution of debris charge $Q_{d}$ in an open-boundary plasma
with equilibrium flow $u_{0}=0.4c_{s}$. The debris charge attains
a net negative charge of $\sim10^{9}e$. The shaded time window is
used for the analysis where the debris charge is in statistical equilibrium.
Right:~The ion phase-space at $t=2\times10^{4}\omega^{-1}_{\textrm{pe}}$,
when $Q_{d}$ is in statistical equilibrium.}
\label{fig:Left:Evolution-of-debris}
\end{figure}

\begin{figure}[t]
\begin{centering}
\includegraphics[width=0.495\textwidth]{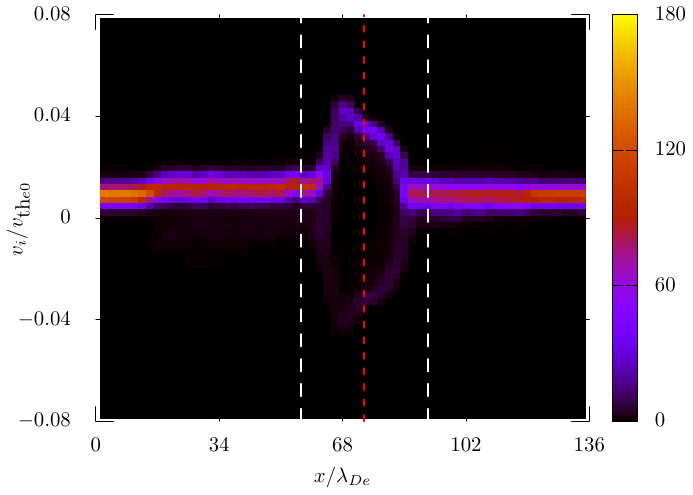}\hfill{}\includegraphics[width=0.495\textwidth]{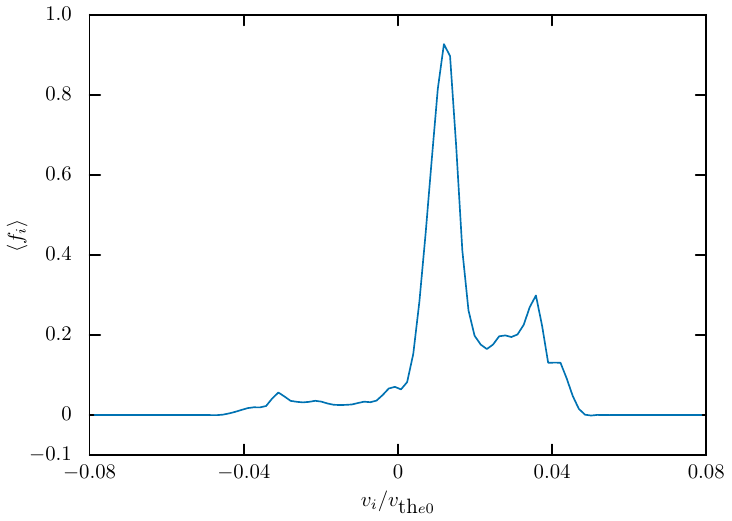}\\
 \includegraphics[width=0.495\textwidth]{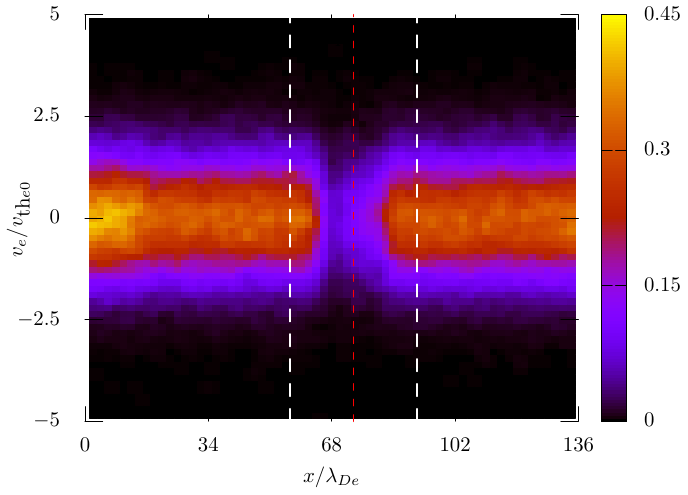}\hfill{}\includegraphics[width=0.495\textwidth]{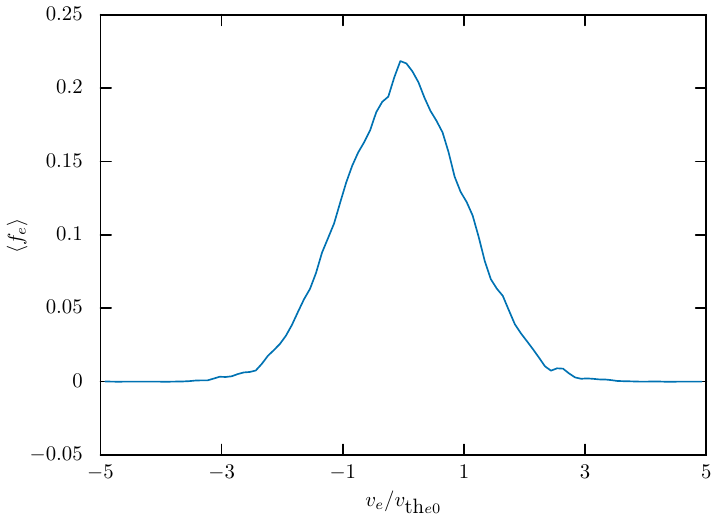} 
\par\end{centering}
\caption{Ion and electron velocity distribution functions $f_{i,e}$ at the
debris site. The regions marked by the white dashed lines with center
marked by the red dashed lines are the regions in which the corresponding
distribution functions are sampled.}
\label{fig:Ion-and-electron} 
\end{figure}

We now examine the kinetic response of an initially uncharged object
embedded in a flowing, collisionless $e$--$i$ plasma. Preferential
electron collection drives the object to a negative potential and
produces the counter-streaming state discussed above, as indicated
by the charge evolution in the left panel of Fig.~\ref{fig:Left:Evolution-of-debris}.
The ion phase-space clearly shows the sustained phase-space vortices
throughout the duration of the simulation (right panel of Fig. \ref{fig:Left:Evolution-of-debris}).
As far as the evolved distributions of electrons and ions go, the
electrons remain close to a Maxwellian, while the ion develops a double-hump
distribution signaling a phase-space resonance phenomenon at the site
of the debris. In Fig. \ref{fig:Ion-and-electron}, we show the velocity
distribution functions $f_{i,e}$ at the debris site. In the figure,
the regions marked by the white vertical dashed lines with the center
marked by the red dashed line in the heat maps are the regions which
are used for sampling the respective distribution functions. Away
from the perturbation, both electron and ion distributions remain
Maxwellian as expected. The observed charge modulation, ion phase-space
structure, and double-hump ion distribution are consistent with an
IICSI-related kinetic response.

\section{Precursor and wake diagnostics}\label{sec:Precursor-and-wake}

We have configured the \emph{h}-PIC-MCC code to auto-detect the precursor
and wake regions based on a controlled simulation run without any
debris, with the exact plasma conditions, including equilibrium plasma
flow. This controlled case supplies us with background fluctuations
against which the wake and precursor regions are detected.

\subsection{Preparing the diagnostics}

To start with, in the diagnostic regions (the precursor and wake regions),
we smoothen the required plasma parameters through a binomial spatiotemporal
smoother 
\begin{eqnarray}
g^{(x)}_{j,n} & \leftarrow & \frac{g_{j-1,n}+2g_{j,n}+g_{j+1,n}}{4},\\
g^{(t)}_{j,n} & \leftarrow & \frac{g_{j,n-1}+2g_{j,n}+g_{j,n+1}}{4},
\end{eqnarray}
where 
\begin{equation}
g_{j,n}=g(x_{j},t_{n})\in\{n_{i},n_{e},u_{i},u_{e},\phi,E\}
\end{equation}
is the respective plasma field at spatial point $x_{j}$ and at the
saved time snapshot $t_{n}$.

To determine the diagnostics regions, we consider the perturbed plasma
field $\tilde{g}_{d}$ due to the debris, which is the plasma field
after corresponding background subtraction. For any field $g$, the
corresponding background temporal mean $\mu_{g,\textrm{control}}$
is determined from the controlled run 
\begin{equation}
\mu_{g,\textrm{control}}(x)=\left\langle g_{\textrm{control}}(x,t)\right\rangle _{t\in\mathcal{T}},\label{eq:temporal_mean-1}
\end{equation}
where $\left\langle \cdot\right\rangle $ indicates temporal average
\begin{equation}
\left\langle A\right\rangle =\frac{1}{\Delta T}\int_{{\cal T}}A(x,t)\,dt,\quad\Delta T=t_{2}-t_{1},\label{eq:temporal_mean-2}
\end{equation}
and $\mathcal{T}=[t_{1},t_{2}]$ is its selected temporal interval.
The debris perturbation fields are then given by 
\begin{align}
\widetilde{n}_{i,d}(x,t) & =\frac{n_{i,d}(x,t)-\mu_{n_{i}\textrm{control}}(x)}{\mu_{n_{i}\textrm{control}}(x)}, & \widetilde{n}_{e,d}(x,t) & =\frac{n_{e,d}(x,t)-\mu_{n_{e},\textrm{control}}(x)}{\mu_{n_{e},\textrm{control}}(x)},\label{eq:control1}\\
\widetilde{u}_{i,d}(x,t) & =\frac{u_{i,d}(x,t)-\mu_{u_{i},\textrm{control}}(x)}{c_{s}}, & \widetilde{u}_{e,d}(x,t) & =\frac{u_{e,d}(x,t)-\mu_{u_{e},\textrm{control}}(x)}{c_{s}},\\
\widetilde{\phi}_{d}(x,t) & =\phi_{d}(x,t)-\mu_{\phi,\textrm{control}}(x), & \widetilde{E}_{d}(x,t) & =E_{d}(x,t)-\mu_{E,\textrm{control}}(x),\label{eq:control3}
\end{align}
where 
\begin{equation}
c_{s}=\frac{1}{\sqrt{m_{i}/m_{e}}}.
\end{equation}
The division in case of perturbed densities by $\mu_{g,\textrm{control}}$
converts it to a fractional density perturbation 
\begin{equation}
\widetilde{n}_{d}\sim\frac{\delta n}{n_{0}}.
\end{equation}
On the other hand, the division by $c_{s}$ for velocity perturbation
measures the perturbed velocity relative to the local Mach-number
\begin{equation}
\widetilde{u}_{d}\sim\frac{\delta u}{c_{s}}.
\end{equation}
We also construct temporal fluctuation of the plasma fields 
\begin{equation}
\delta_{t}g(x,t)=g(x,t)-\langle g(x,t)\rangle_{t}.\label{eq:debris-fluct}
\end{equation}
for spectral diagnostics. This is necessary as $\tilde{g}_{d}$ retains
both stationary and time-dependent debris-induced changes relative
to the controlled environment, whereas $\delta_{t}g_{d}$ removes
the debris-induced spatial structure and retains only its temporal
part.

\subsection{Auto-detection of diagnostic regions}

All throughout our discussion, we shall assume a right-moving equilibrium
plasma flow, so that the precursor region lies left of the debris
and wake extends to the right of the debris field. The sequence is
\begin{equation}
\ensuremath{\text{inflow}\longrightarrow\text{upstream plasma}\longrightarrow\text{debris}\longrightarrow\text{downstream wake}.}
\end{equation}

In order to determine the extent of the diagnostic regions, an immediate
sheath half-width $\ell_{s}=\epsilon_{s}\lambda_{De}$ is excluded
from the vicinity of the debris, where at present we have set $\epsilon_{s}=3$.
The reservoir buffer and boundary zones are set to a width of $20\lambda_{De}$
at each end, and are excluded from the diagnostic regions. Therefore,
precursor and wake points satisfy the following 
\begin{align}
L_{\textrm{buffer}} & \leq x\leq x_{d}-\ell_{s},\\
x_{d}+\ell_{s} & \leq x\leq L-L_{\textrm{buffer}},
\end{align}
respectively. The \emph{h}-PIC-MCC code uses the normalized fields
$(\widetilde{n}_{i},\widetilde{u}_{i},\widetilde{\phi},\widetilde{E})$
for auto detection. Each one of these fields is measured relative
to the local time-averaged control value (from the corresponding controlled
run) at the same $x$. Consequently, precursor and wake detection
does not require a supposedly quiet fixed upstream or downstream interval.
For each field $g$, the selected snapshots are divided into blocks.
The configured block length is currently set to $N_{\textrm{block}}=40$
snapshots, although this number is arbitrary. A shorter dataset is
divided into at least two blocks whenever possible. Let $B$ be the
number of resulting blocks. For a certain block, the local mean-square
power is calculated through 
\begin{equation}
P_{g,\textrm{block}}(x)=\left\langle |\widetilde{g}(x,t)|^{2}\right\rangle _{t\in\textrm{block}}.
\end{equation}
The analyzer then calculates the block mean $\overline{P}_{g,d}(x)$
\begin{equation}
\overline{P}_{g,d}(x)=\frac{1}{B}\sum^{B}_{\textrm{block}=1}P_{g,\textrm{block}}(x)
\end{equation}
and its standard error for both debris and control data. The spatial
power profiles are smoothed with the same binomial operator, as mentioned
before. For a position in a block to be designated inside a diagnostic
region (either precursor or wake), at least two of the fields $(\widetilde{n}_{i},\widetilde{u}_{i},\widetilde{\phi},\widetilde{E})$
must satisfy the following two conditions 
\begin{align}
\mathcal{R}_{g}(x) & =\frac{\overline{P}_{g,d}(x)}{\max[\overline{P}_{g,\textrm{control}}(x),\epsilon_{\textrm{floor}}]}\geq\mathcal{R}_{\min},\label{eq:cond1}\\
Z_{g}(x) & =\frac{\overline{P}_{g,d}(x)-\overline{P}_{g,\textrm{control}}(x)}{\sqrt{\sigma^{2}_{g,d}(x)+\sigma^{2}_{g,\textrm{control}}(x)+[10^{-6}P_{\mathrm{scale}}]^{2}}}\geq Z_{\min},\label{eq:cond2}
\end{align}
where $\sigma_{g,d}$ is the respective standard error for a diagnostic
block (or an equivalent controlled snapshot), derived from the respective
standard deviation $s_{g,d}$, 
\begin{eqnarray}
s_{g,d}(x) & = & \sqrt{\frac{1}{B-1}\sum^{B}_{\textrm{block}=1}\left[P_{g,\textrm{block}}(x)-\overline{P}_{g,d}(x)\right]^{2}},\\
\sigma_{g,d}(x) & = & \frac{s_{g,d}(x)}{\sqrt{B}}.
\end{eqnarray}
For the controlled run also, we use the same corresponding expressions.
The first condition (\ref{eq:cond1}) measures the power-ratio for
a field ${\cal R}_{g}(x)$ and it must stay at least ${\cal R}_{\textrm{min}}$
times above the control power. Simultaneously, the second condition
(\ref{eq:cond2}) measures the power difference $Z_{g}(x)$ for a
field between a diagnostic block and a control block and it must stay
at least $Z_{\textrm{min}}$ number of standard errors above zero.
Presently, we set these values as 
\begin{equation}
\mathcal{R}_{\min}=1.5,\qquad Z_{\min}=3.
\end{equation}
which translates to the conditions that for a block to qualify as
a diagnostic block, its fluctuation power (debris-induced perturbation)
must be at least $1.5$ times the control power and the power difference
must be at least $3$ times the combined errors above zero. The small
number $\epsilon_{\textrm{floor}}$ is introduced to prevent a division
by zero in case of an unrealistically small denominator. The factor
$[10^{-6}P_{\mathrm{scale}}]$ also serves a similar purpose for $Z_{g}(x)$,
where $P_{\mathrm{scale}}=\max_{x}\overline{P}_{g,\textrm{control}}(x)$.

With the above checks in place, the algorithm searches away from the
debris in both directions, starting at the outer edges of the excluded
debris sheath. The search continues to the last significant point
and it stops after a certain number of contiguous non-significant
regions are found. Currently, we stop searching for diagnostic region
after $8$ contiguous non-significant spatial bins are detected.

\subsection{Auto-probing}

Once the diagnostic regions have been identified, numerical probes
are used to sample the perturbations on both sides of the debris.

The analyzer performs an automatic moving-probe scan rather than using
only one permanently fixed set of absolute positions. This is performed
in two steps. At the first step, the diagnostic probe is fixed at
$x_{r}$, relative to the configured debris position 
\begin{equation}
x_{r}=x_{d}-5\lambda_{De},
\end{equation}
and the probe position $x_{p}$ is set with 
\begin{equation}
x_{p}=x_{d}+\Delta x_{p},\quad\Delta x_{p}\in\{-30,-20,-10,10,20,30\}\lambda_{De}.
\end{equation}
This tests whether measured delays are consistent with a source at
the known $x_{d}$. In the second stage, the actual blind localization
automatically scans trial source positions $x_{\textrm{trial}}$ and
for every trial position, it relocates the complete probe set as per
\begin{eqnarray}
x_{r}(x_{\textrm{trial}}) & = & x_{\textrm{trial}}-5\lambda_{De},\\
x_{p}(x_{\textrm{trial}}) & = & x_{\textrm{trial}}+\Delta x_{p}.
\end{eqnarray}
Thus, if the trial position changes, all reference and secondary probes
move with it. At every trial position, the algorithm 
\begin{enumerate}
\item extracts the field time series at the trial-relative probe positions, 
\item calculates cross-correlations, 
\item scans the automatically selected perturbation velocities (see next
section), 
\item computes the delay-and-sum score (see next section), and 
\item moves to the next trial position. 
\end{enumerate}

\subsection{Mapping the perturbation velocity}

In order to detect the debris position, we need to know the propagation
velocity of the debris-induced perturbation (IICSI) across the diagnostic
regions. As the simulation propagates, it records the time evolution
of the perturbation.

The velocity of propagation of perturbation $v_{\textrm{pert}}$ can
be expressed as 
\begin{equation}
v_{\textrm{pert}}=u_{0}+v_{\textrm{rel}},
\end{equation}
where $u_{0}$ is the equilibrium plasma flow and $v_{\textrm{rel}}$
is the propagation velocity of the perturbation measured in the rest
frame of the plasma. For ion-acoustic-like disturbance, we have 
\begin{equation}
v_{\textrm{rel}}\approx\pm c_{s},
\end{equation}
and we have thus two lab-frame branches 
\begin{equation}
v_{\textrm{pert}\pm}\approx u_{0}\pm c_{s}.
\end{equation}
Writing the equilibrium flow as $u_{0}=Mc_{s}$, we have 
\begin{equation}
\frac{v_{\textrm{pert}\pm}}{c_{s}}\approx M\pm1.
\end{equation}
Thus the analyzer does not force the measured perturbation to follow
exactly $u_{0}\pm c_{s}$, rather it scans a range of trial velocities
as the actual propagation speed of the perturbation can be altered
by dispersion, kinetic effects, trapping, nonlinear structures, local
temperature changes, and the debris sheath. In particular, we define
a scan limit 
\begin{equation}
V_{\max}/c_{s}=\max(2,|M|+1).
\end{equation}

\begin{figure}
\begin{center}\small
\begin{tikzpicture}[
    node distance=7mm and 8mm,
    flow/.style={
        draw,
        rounded corners=2mm,
        fill=blue!6,
        inner xsep=7pt,
        inner ysep=4pt,
        font=\small
    },
    arrow/.style={
        -{Stealth[length=2mm,width=1.5mm]},
        thick
    }
]

% First row
\node[flow] (a) {Field preparation};
\node[flow, right=of a] (b) {Precursor/wake detection};
\node[flow, right=of b] (c) {Probe placement};

% Second row
\node[flow, below=7mm of c] (d) {Delay analysis};
\node[flow, left=of d] (e) {Source scan};
\node[flow, left=of e] (f) {Field detection};

% First-row arrows
\draw[arrow] (a) -- (b);
\draw[arrow] (b) -- (c);

% Compact return arrow
\draw[arrow, rounded corners=2.5mm]
    (c.east) -- ++(5mm,0) |- (d.east);

% Second-row arrows
\draw[arrow] (d) -- (e);
\draw[arrow] (e) -- (f);

\end{tikzpicture}
\end{center}

\caption{Flow chart for estimation of debris position.}\label{fig:Flow-chart-for}
\end{figure}
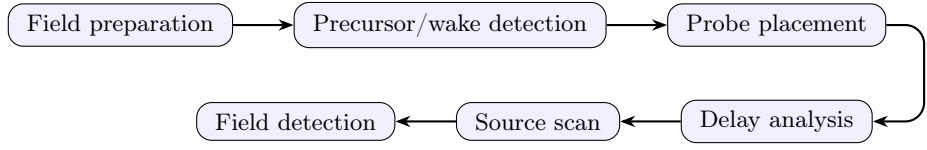

\section{Propagation-delay framework}\label{sec:Propagation-delay-framework}

We now examine whether the probe diagnostics can provide information
about the source position through measurements of different plasma
fields. This section describes the propagation-delay procedure in
detail. Its purpose is to establish whether the precursor and wake
retain a source-centered temporal organization that can support localization.
It should be noted that this is not presented as an operational detection
procedure. The analysis follows the sequence shown in Fig.~\ref{fig:Flow-chart-for}.
The plasma fields are first reconstructed and smoothed, and the matching
control background is subtracted to isolate the debris-induced perturbations.
Numerical probes then sample these perturbations on both sides of
each trial source position. The corresponding probe-delay correlations
and a flow-dependent propagation-velocity scan are used to test the
trial positions, after which the scores from $n_{i}$, $u_{i}$, $\phi$,
and $E$ are combined.

\subsection{Probe-delay diagnostics}

Before calculating a probe correlation, we perform a second subtraction,
\begin{equation}
\ensuremath{s_{x}(t)=\widetilde{g}_{d}(x,t)-\left\langle \widetilde{g}_{d}(x,t)\right\rangle _{t}}.
\end{equation}
This is also carried out for the corresponding control-run signal.
The correlation is now defined as 
\begin{equation}
\rho_{rp}(\tau)=\frac{\sum_{t}s_{r}(t)\,s_{p}(t+\tau)}{\sqrt{\sum_{t}s^{2}_{r}(t)\,\sum_{t}s^{2}_{p}(t+\tau)}},
\end{equation}
where $(r,p)$ stand for reference probe and secondary probe, and
$\tau$ is the time lag between their signals. The analyzer searches
a configured lag interval $|\tau|\leq1000\,\omega^{-1}_{\textrm{pe}}$,
and retains the lag at the largest $|\rho|$. It then fits 
\begin{equation}
\tau_{p}=a+\frac{x_{p}-x_{r}}{v_{\mathrm{fit}}},
\end{equation}
by least-squares fitting. Here $\tau_{p}$ is the measured delay between
reference and secondary probes and $v_{\mathrm{fit}}$ is an empirical
propagation velocity of the perturbation obtained from the measured
delays between several spatial probes. Assume that the reference probe
is at $x_{r}$, and secondary probe $p$ is at $x_{p}$. We then define
its signed separation as 
\begin{equation}
\ensuremath{d_{p}=x_{p}-x_{r}.}
\end{equation}
The measured travel time $\tau^{\textrm{meas}}_{p}$ itself is then
found from the cross-correlation peak 
\begin{equation}
\ensuremath{\tau^{\textrm{meas}}_{p}=\operatorname*{arg\,max}_{\tau}|\rho_{rp}(\tau)|},
\end{equation}
which will produce several measured pairs 
\begin{equation}
\ensuremath{(d_{1},\tau^{\textrm{meas}}_{1}),(d_{2},\tau^{\textrm{meas}}_{2}),\dots,(d_{P},\tau^{\textrm{meas}}_{P})}.
\end{equation}
If one disturbance propagates through all probes with an approximately
constant velocity, the measured delays should satisfy 
\begin{equation}
\ensuremath{\tau^{\textrm{meas}}_{p}\simeq a+\frac{d_{p}}{v_{\textrm{fit}}}.}
\end{equation}

\subsection{Source-centered localization}

The localization stage repeats the delay test for many trial probe
positions $x_{\textrm{trial}}$. We should also note that the probe
positions must leave room for the boundary buffer zones and the largest
probe offset 
\begin{equation}
L_{\textrm{buffer}}+\max|\Delta x_{p}|\leq x_{\textrm{trial}}\leq L-L_{\textrm{buffer}}-\max|\Delta x_{p}|.
\end{equation}
For every trial perturbation source position (the debris position),
the reference probe and all secondary probes are moved together, while
their prescribed displacements from the trial position remain unchanged.
For each trial velocity $v_{q}$, the predicted lag for probe $p$
is then given by 
\begin{equation}
\tau_{p,q}=\frac{x_{p}(x_{\textrm{trial}})-x_{r}(x_{\textrm{trial}})}{v_{q}}.
\end{equation}
We then calculate the field-specific delay-and-sum score 
\begin{equation}
S_{g}(x_{\textrm{trial}},v_{q})=\sum_{p}\rho^{2}_{rp}(\tau_{p,q}),
\end{equation}
and 
\begin{equation}
S_{g}(x_{\textrm{trial}})=\max_{v_{q}}S_{g}(x_{\textrm{trial}},v_{q}).
\end{equation}
The source estimate for field $g$ is the detected position that maximizes
$S_{g}(x_{\textrm{trial}})$, so only the maximizing velocity is retained.

\section{Weak-form residual analysis}\label{sec:Weak-form-residual-analysis}

We now use the PIC simulation data for weak-form of sparse identification
(using SINDy \citep{Brunton2016}), which is used here only to obtain
compact bulk equations and their residuals from the PIC fields. Coefficients
are learned from the matched control run and then held fixed when
the same equations are evaluated on the debris run. The resulting
excess residuals show where the debris produces a departure from the
unperturbed bulk dynamics.

Toward this, we define the state to be discovered as 
\begin{equation}
\bm{y}(x,t)=\begin{bmatrix}\td n_{i,d} & \td n_{e,d} & \td u_{i,d} & \td\phi_{d} & \td E_{d}\end{bmatrix}^{\mathsf{T}},
\end{equation}
where the $\,\widetilde{\cdot}\,$ quantities are the controlled-background-subtracted
fields, defined in Eqs. (\ref{eq:control1}-\ref{eq:control3}). Occasionally,
we also use a smaller state such as $\bm{y}=(\td n_{i,d},\td u_{i,d},\td\phi_{d},\td E_{d})^{\mathsf{T}}$,
which is often preferable when an unnecessarily large state increases
collinearity. The desired local model is this 
\begin{equation}
\frac{\partial\bm{y}}{\partial t}=\bm{\mathcal{F}}\left(\bm{y},\partial_{x}\bm{y},\partial^{2}_{x}\bm{y},\partial^{3}_{x}\bm{y},\ldots\right)+\bm{r},
\end{equation}
where $\bm{\mathcal{F}}$ is an unknown function and $\bm{r}$ contains
residuals such as particle noise, derivative error, unresolved kinetic
physics, and model discrepancy.

As mentioned above, in this analysis, we use the integral form of
the SINDy algorithm rather than the direct one, which is better suited
to noisy field data because the weak formulation avoids point-wise
differentiation \citep{Messenger2021a,Messenger2021b}.

\subsection{Discovery region and data preparation}

From every data frame, the code reconstructs the one-dimensional fields
\begin{equation}
n_{i}(x,t),\qquad u_{i}(x,t),\qquad p_{i}(x,t),\qquad\phi(x,t),\qquad E(x,t).
\end{equation}
Each field is smoothed along the spatial direction by binomial passes.
We consider a so-called ``weak spatial window'', which is a small
spatial interval of the simulation domain over which the PDE is integrated
instead of being enforced at one grid point. For such a window centered
at grid point $x_{j}$, we form the domain as 
\begin{equation}
\mathcal{W}^{x}_{j}=[x_{j-h_{x}},x_{j+h_{x}}],
\end{equation}
where $h_{x}$ is the spatial half-width, which is set to $3$ in
the present case. Thus, provided the center point is included, each
spatial window contains $2h_{x}+1=7$ numbers of grid points. Inside
this interval, the code applies the smooth weight 
\begin{equation}
w_{x}(x)=\sin^{2}\!\left[\pi\left(\frac{x-x_{L}}{x_{R}-x_{L}}\right)\right],
\end{equation}
where $x_{L}$ and $x_{R}$ are the left and right window boundaries.
The weight is zero at both boundaries and largest near the center.

For example, the point-wise continuity equation 
\begin{equation}
\partial_{t}n_{i}+\partial_{x}(n_{i}u_{i})=0,
\end{equation}
is multiplied by $w_{x}(x)$ and integrated over the window. Integration
by parts results in 
\[
\int^{x_{R}}_{x_{L}}w_{x}\,\partial_{x}(n_{i}u_{i})\,dx=-\int^{x_{R}}_{x_{L}}w_{x}'(x)n_{i}u_{i}\,dx.
\]
In the above expression, the first part (of the integration by parts)
contains the boundary terms, which vanish because $w_{x}(x_{L})=w_{x}(x_{R})=0$.
This way we avoid calculating the noisy derivative $\partial_{x}(n_{i}u_{i})$
at the center point. It integrates the measured flux $(n_{i}u_{i})$
against the exactly known derivative $w_{x}'$. Combined with the
temporal window, the complete weak space-time window is given by 
\begin{equation}
\mathcal{W}_{j,a}=\left[x_{j-h_{x}},x_{j+h_{x}}\right]\times\left[t_{a},t_{a+N_{w}-1}\right],
\end{equation}
where $N_{w}$ is the number of snapshots in each time window, which
is set to $200$ for the localized weak-residual calculation, with
a stride of $20$ snapshots. In the above expression, $a$ labels
the starting snapshot of the temporal window. The more general integral
discovery scan uses a $400$-snapshot window and a stride of $40$
snapshots.

For discovery, we only use the physically admissible diagnostic regions
$\Omega$ i.e.\ the designated precursor and wake 
\begin{equation}
\Omega_{\mathrm{pre}}=\{x:L_{\mathrm{buffer}}\le x\le x_{d}-\ell_{s}\},\qquad\Omega_{\mathrm{wake}}=\{x:x_{d}+\ell_{s}\le x\le L-L_{\mathrm{buffer}}\}.
\end{equation}
This excludes the open-boundary buffers and the immediate debris sheath,
where particle absorption and the localized charge $\rho_{d}$ cannot
generally be represented by a homogeneous fluid closure. Mixing these
regions with the remote precursor or wake could introduce spurious
higher-order terms; the two remote regions are therefore fitted separately,
\begin{equation}
\bm{\Xi}_{\mathrm{pre}}\ne\bm{\Xi}_{\mathrm{wake}},
\end{equation}
where $\Xi$ is the SINDy coefficient matrix. In general, these matrices
are unequal unless there is no or minimal equilibrium plasma flow.

For a general regional fit, the retained dataset can be written as
\begin{equation}
\mathcal{D}=\{(x_{j},t_{n}):x_{j}\in\Omega_{\mathrm{fit}},\ t_{n}\in\mathcal{T}_{\mathrm{fit}}\},
\end{equation}
where $\Omega_{\mathrm{fit}}$ and $\mathcal{T}_{\mathrm{fit}}$ are
the spatial region used for fitting and the selected fitting-time
interval, respectively. The interval $\mathcal{T}_{\mathrm{fit}}$
is selected only after the debris charge reaches a statistical equilibrium
state.

The numerical parameters used in the regression and localization analysis
are given in Table. (Appendix C). The debris run is analysed after
$t=10^{4}\omega^{-1}_{\mathrm{pe}}$, when its charge has reached
a statistical equilibrium as shown in Fig. \ref{fig:Left:Evolution-of-debris}.
Accordingly, the same time interval is used for the matched control
run as well. The simulations were performed using a multicore-parallel
Julia implementation.

\subsection{Weak formulation}

The weak formulation is preferable for PIC data because direct numerical
differentiation amplifies particle noise. A governing equation is
multiplied by a smooth test function and integrated over a local space--time
window; integration by parts transfers derivatives from the measured
field to the known test function \citep{Messenger2021a,Messenger2021b}.

In our case, we define a spatiotemporal coordinate $z\in[z_{1},z_{2}]$
(which represents either spatial or temporal coordinate) and a smooth
test function 
\begin{equation}
w(z)=\sin^{2}\left(\pi\frac{z-z_{1}}{z_{2}-z_{1}}\right).
\end{equation}
Its first two derivatives, evaluated analytically by the code, are
\begin{align}
w'(z) & =\frac{\pi}{z_{2}-z_{1}}\sin\left(2\pi\frac{z-z_{1}}{z_{2}-z_{1}}\right),\\
w''(z) & =\frac{2\pi^{2}}{(z_{2}-z_{1})^{2}}\cos\left(2\pi\frac{z-z_{1}}{z_{2}-z_{1}}\right).
\end{align}
We use separate functions $w_{x}(x)$ and $w_{t}(t)$ for both space
and time coordinate in each weak space-time window. Because they vanish
at the endpoints, integration by parts moves the principal time and
space derivatives from the noisy PIC fields onto known smooth functions.
For a field $g$, the weak integral is given by 
\begin{equation}
\mathcal{I}[g;A,B]=\int^{t_{R}}_{t_{L}}\int^{x_{R}}_{x_{L}}g(x,t)\,A(x)\,B(t)\,dx\,dt.
\end{equation}
As the PIC code provides values only at discrete positions and times,
we convert the integral to its discrete counterpart 
\begin{equation}
\mathcal{I}[g;A,B]\approx\sum_{r}\sum_{s}g(x_{r},t_{s})\,A(x_{r})\,B(t_{s})\,q_{x,r}\,q_{t,s}.
\end{equation}
The function $A(x)$ is the spatial weighting function and depending
on the PDE term, it can be set to $w_{x}(x)$ or $w_{x}'(x)$ or $w_{x}''(x)$,
etc. Similarly, $B(t)$ is the temporal weighting function and in
the present fluid formalism, it is usually 
\begin{equation}
B(t)=w_{t}(t)\,\textrm{or}\,\ensuremath{w_{t}'(t).}
\end{equation}

For the sampled coordinates 
\begin{equation}
z_{1}<z_{2}<\cdots<z_{N},
\end{equation}
we employ the trapezoidal rule 
\begin{equation}
\int^{z_{N}}_{z_{1}}f(z)\,dz\approx\sum^{N}_{k=1}q_{k}f_{k},
\end{equation}
where $q_{k}$ are the trapezoidal weights 
\begin{equation}
q_{1}=\frac{z_{2}-z_{1}}{2},\qquad q_{N}=\frac{z_{N}-z_{N-1}}{2},\qquad q_{k}=\frac{z_{k+1}-z_{k-1}}{2}.
\end{equation}

\subsection{Control-trained fluid equations}

We now construct the weak forms of the control-trained fluid equations.

\subsubsection{Weak ion-continuity equation}

Away from a localized particle source or sink, the bulk continuity
equation is 
\begin{equation}
\partial_{t}n_{i}+\partial_{x}(n_{i}u_{i})=0.
\end{equation}
Multiplication of the above equation by $w_{x}w_{t}$, followed by
integration over $\mathcal{W}_{j,a}$ results in the weak integral
form of the ion-continuity equation 
\begin{equation}
-\mathcal{I}[n_{i};w_{x},w_{t}']=\xi_{c}\,\mathcal{I}[n_{i}u_{i};w_{x}',w_{t}]+\bm{r}_{c}(j,a).\label{eq:sindyeq}
\end{equation}
The subscript `$c'$ . in the above equation denotes the `continuity'
equation. For a perfect ion-continuity equation without a source or
sink, $\xi_{c}=1$ and $r_{c}=0$. Accordingly, $\xi_{c}$ multiplies
for the conservative ion-transport term in the weak continuity equation
and $r_{c}$ is the corresponding residual which includes noise, derivative
error, unresolved physics, and model discrepancy, etc. In Eq. (\ref{eq:sindyeq}),
$\bm{b}_{c}(j,a)=-\mathcal{I}[n_{i};w_{x},w_{t}']$ is the response
or target vector and $\bm{\Psi}_{c}(j,a)=\mathcal{I}[n_{i}u_{i};w_{x}',w_{t}]$
is the candidate-library matrix (also known as feature matrix), so
that the corresponding regression is 
\begin{equation}
\bm{b}_{c}(j,a)=\bm{\xi}_{c}\bm{\Psi}_{c}(j,a)+\bm{r}_{c}(j,a).\label{eq:regre1}
\end{equation}

\subsubsection{Weak ion-momentum equation}

The configured bulk momentum library contains 
\begin{equation}
-u_{i}\partial_{x}u_{i},\quad-\frac{q_{i}}{m_{i}}\partial_{x}\phi,\quad-\frac{1}{m_{i}n_{i}}\partial_{x}p_{i},\quad\partial^{2}_{x}u_{i},\quad1.
\end{equation}
The corresponding weak response is 
\begin{equation}
b_{m}(j,a)=-\mathcal{I}[u_{i};w_{x},w_{t}'],
\end{equation}
where subscript `$m$' denotes `momentum' equation. The candidate
columns used in the regression analysis are 
\begin{align}
\Psi_{m,1}(j,a) & =\mathcal{I}\!\left[\frac{u^{2}_{i}}{2};w_{x}',w_{t}\right],\\
\Psi_{m,2}(j,a) & =\frac{q_{i}}{m_{i}}\mathcal{I}[\phi;w_{x}',w_{t}],\\
\Psi_{m,3}(j,a) & =\mathcal{I}\!\left[-\frac{\partial_{x}p_{i}}{m_{i}\max(n_{i},\epsilon_{n})};w_{x},w_{t}\right],\\
\Psi_{m,4}(j,a) & =\mathcal{I}[u_{i};w_{x}'',w_{t}],\\
\Psi_{m,5}(j,a) & =\mathcal{I}[1;w_{x},w_{t}].
\end{align}
The first, second, and fourth expressions can be derived from the
integration by parts. The pressure term is an exception, as due the
factor $1/n_{i}$, it is not amenable to integration by parts. So,
we smooth the pressure field $p_{i}$ and evaluate $\partial_{x}p_{i}$
with a centered nonuniform-grid difference in the interior and one-sided
differences at the ends of the local window, and protect the density
denominator by $\max(n_{i},\epsilon_{n})$ from division by zero,
$\epsilon_{n}$ being a very small positive number. Again, the regression
has the matrix form 
\begin{equation}
\bm{b}_{m}(j,a)=\bm{\Psi}_{m}(j,a)\bm{\xi}_{m}+\bm{r}_{m}(j,a).\label{eq:regre2}
\end{equation}

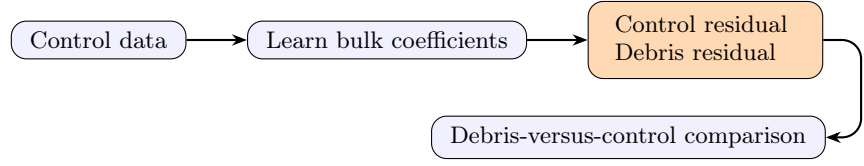
\begin{figure}
\begin{center}\small
\begin{tikzpicture}[
    node distance=7mm and 8mm,
    flow/.style={
        draw,
        rounded corners=2mm,
        fill=blue!6,
        inner xsep=7pt,
        inner ysep=4pt,
        font=\small
    },
    arrow/.style={
        -{Stealth[length=2mm,width=1.5mm]},
        thick
    }
]

% First row
\node[flow] (a) {Control data};

\node[flow, right=of a] (b)
    {Learn bulk coefficients};

\node[flow, right=of b, align=left,fill=orange!30] (c) {%
    $
    \begin{array}{l}
        \text{Control residual}\\
        \text{Debris residual}
    \end{array}$
};

% Second row
\node[flow, below=7mm of c] (d) at (7,-.3)
    {Debris-versus-control comparison};

% First-row arrows
\draw[arrow] (a) -- (b);
\draw[arrow] (b) -- (c);

% Bent return arrow
\draw[arrow, rounded corners=2.5mm]
    (c.east) -- ++(5mm,0) |- (d.east);

\end{tikzpicture}
\end{center}

\caption{The regression sequence for weak SINDy for debris localization.}\label{fig:The-regression-sequence}
\end{figure}

Both Eqs. (\ref{eq:regre1}) and (\ref{eq:regre2}) are used for the
control-run as well as for the run with debris which provide the debris-versus-control
comparison. This sequence is shown in Fig.\ref{fig:The-regression-sequence}.

\subsubsection{Sparse regression and residual maps}

For sparse regression, the normal bulk-ion equation is learned from
the control-run, which is an exact replica of the required simulation
without the debris. In the code, we use sequentially thresholded ridge
least squares (STLSQ) for stabilization, following the sparse-regression
strategy introduced with SINDy \citep{Brunton2016}. The code is designed
to choose an equation that captures the ordinary plasma dynamics in
the no-debris control run while retaining only the necessary physical
terms. The coefficients of this equation are determined from the control
run and then used without change for both the control and debris runs.
Consequently, any additional effect produced near the debris remains
visible in the residual instead of being incorrectly represented as
a change in the ordinary plasma transport or force terms.

For the continuity and momentum equations $(c,m)$, the control-trained
weak residual is given by 
\begin{equation}
r_{(c,m)}(j,a)=b_{(c,m)}(j,a)-\sum_{k}\widehat{\xi}_{(c,m),k}\Psi_{(c,m),k}(j,a),
\end{equation}
where $\widehat{\bm{\xi}}$ is the numerical estimate of $\xi$, produced
by weak regression using the control run. It produces two residuals,
namely $r^{(d)}$ for the debris-run and $r^{(0)}$ for the control-run.
While the latter measures the residual normally generated by PIC noise,
discretization, smoothing, imperfect fluid closure, and background
kinetic fluctuations, the former contains these contributions plus
any additional localized source or sink represented by the debris.
A debris signature is therefore not merely a large residual, it must
be spatially localized and exceed the matched control background.

The residual map is projected onto assumed localized Gaussian source
functions $G_{\ell}$, centered at several trial positions $x_{\ell}$
\begin{equation}
G_{\ell}(x)\propto\exp\left[-\frac{(x-x_{\ell})^{2}}{2\sigma^{2}_{G}}\right],
\end{equation}
where $\sigma_{G}=3\lambda_{De}$ is the spatial width of the trial
source function. The residual is projected onto $25$ trial centers,
using a source ridge parameter of $10^{-8}$. We approximate the residual
in a temporal window $a$ as 
\begin{equation}
r_{(c,m)}(x_{j},a)\approx\sum_{\ell}A_{a\ell}G_{\ell}(x_{j}),
\end{equation}
were $A_{a\ell}$ is the fitted amplitude of Gaussian source $\ell$
during time window $a$. We then combine the amplitudes from all time
windows using 
\begin{equation}
S_{\ell}=\sqrt{\frac{1}{N_{\tau}}\sum^{N_{\tau}}_{a=1}A^{2}_{a\ell}}
\end{equation}
with $N_{\tau}$ being the number of temporal windows. The quantity
$S_{\ell}$ is also known as the persistent root-mean-square residual
amplitude associated with each trial source. The procedure is performed
independently for the debris and control runs. The possible debris
location is the center with the largest robust debris-over-control
excess, 
\begin{equation}
\widehat{x}_{d}=x_{\ell_{*}},\qquad\ell_{*}=\argmaxop_{\ell}Z_{\ell}.
\end{equation}
It is accepted only when $Z_{\ell}$ is above a threshold value and
the ratio of $S_{\ell}$ for detected debris to that for the control
run also above a threshold value 
\begin{equation}
Z_{\ell_{*}}\ge Z_{\textrm{thr}},\qquad\frac{S^{(d)}_{\ell_{*}}}{S^{(0)}_{\ell_{*}}+\epsilon_{s}}\ge S_{\textrm{thr}}.
\end{equation}
In the present simulation $Z_{\textrm{thr}}$ is set to $5$ and $S_{\textrm{thr}}$
is set to $2$, and $\epsilon_{s}$ is a small number introduced to
prevent division by zero. In the above expressions 
\begin{equation}
Z_{\ell}=\frac{S^{(d)}_{\ell}-S^{(0)}_{\ell}}{\sigma_{0}},
\end{equation}
where $\sigma_{0}$ is the Gaussian consistency factor.

\begin{table}
\centering{}\caption{Poisson-residual localization. Positions and errors are in $\lambda_{D}$.\vrr}\label{tab:Poisson-residual-localization.}
\begin{tabular}{cccc}
\toprule 
$M$  & configured $x_{d}$  & calculated $\widehat{x}_{d}$  & $\widehat{x}_{d}-x_{d}$\vrr\tabularnewline
\midrule 
0.0  & 68  & 67.9964  & $-0.0036$\vrr\tabularnewline
0.4  & 68  & 67.9967  & $-0.0033$\vrr\tabularnewline
1.0  & 90  & 90.1609  & $+0.1609$\vrr\tabularnewline
1.5  & 112  & 112.5191  & $+0.5191$\vrr\tabularnewline
2.0  & 90  & 90.8465  & $+0.8465$\vrr\tabularnewline
3.0  & 70  & 68.8897  & $-1.1103$\vrd\tabularnewline
\bottomrule
\end{tabular}
\end{table}

\subsection{Vlasov residual localization}

We also evaluate the normalized collisionless ion Vlasov equation
in the weak form. Its residual provides an independent kinetic diagnostic,
\begin{equation}
\partial_{t}f_{i}=-v\,\partial_{x}f_{i}-\frac{q_{i}}{m_{i}}E\,\partial_{v}f_{i},
\end{equation}
where $f_{i}$ is the ion velocity distribution. Over a time window
$[t_{a},t_{b}]$, it constructs the kinetic residual 
\begin{equation}
r_{f}(x,v;[t_{a},t_{b}])=\bar{f}_{i}(x,v,t_{b})-\bar{f}_{i}(x,v,t_{a})-\int^{t_{b}}_{t_{a}}\left(-v\partial_{x}\bar{f}_{i}-\frac{q_{i}}{m_{i}}E\,\partial_{v}\bar{f}_{i}\right)dt,
\end{equation}
where $\bar{f}_{i}$ is a spatially smoothed distribution function.
One smoothing pass is applied, four spatial bins and two velocity
bins are trimmed from the respective boundaries, and non-overlapping
$40$-snapshot windows are used. The time integrals use trapezoidal
quadrature. The simulation forms three velocity moments of the residual
$r_{f}$, 
\begin{align}
r_{N}(x;[t_{a},t_{b}]) & =\int r_{f}\,dv,\\
r_{P}(x;[t_{a},t_{b}]) & =\int vr_{f}\,dv,\\
r_{K}(x;[t_{a},t_{b}]) & =\int\frac{v^{2}}{2}r_{f}\,dv,
\end{align}
representing particle $(N)$, momentum $(P)$, and kinetic energy
$(K)$ residuals. Each moment is localized with the same Gaussian-group
and debris-versus-control comparison described before.

Combining the fluid residuals with the Vlasov residuals can help us
predict the debris location from the precursor and wake dynamics.

\section{Numerical results}\label{sec:Numerical-results}

We first verify the field reconstruction through the Poisson residual
and then characterize the spatial and spectral structure of the precursor
and wake. The final part addresses the inverse problem using measurements
away from the immediate debris region.

\subsection{Poisson residuals}

\begin{figure}
\begin{centering}
\includegraphics[width=0.5\textwidth]{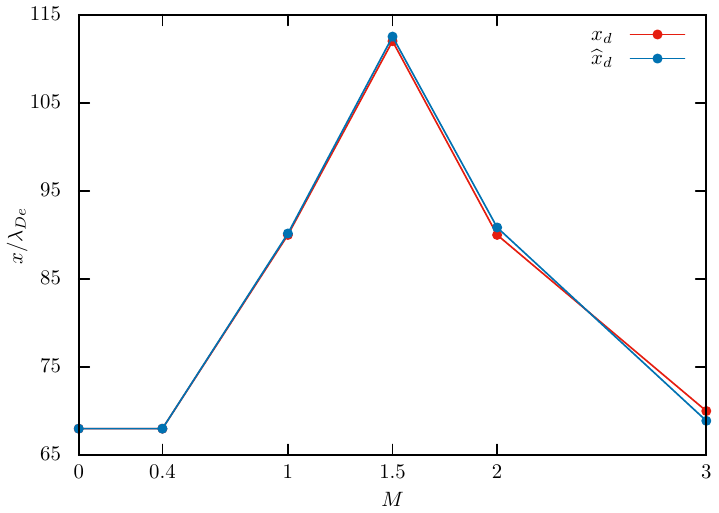} 
\par\end{centering}
\caption{Configured debris position $x_{d}$ and the position $\widehat{x}_{d}$
recovered in the Poisson-residual consistency check.}
\label{fig:Configured-debris-position} 
\end{figure}

As a consistency check, we calculate the Poisson residuals $r_{E,\phi}$,
\begin{align}
r_{E}(x,t) & =\frac{\partial E}{\partial x}-\left(n_{i}-n_{e}\right),\\
r_{\phi}(x,t) & =-\frac{\partial^{2}\phi}{\partial x^{2}}-\left(n_{i}-n_{e}\right),
\end{align}
where $n_{i,e}$ are reconstructed \emph{in situ} from the simulated
distribution functions, 
\begin{equation}
n_{s}(x,t)=\int f_{s}(x,v,t)\,dv.
\end{equation}
Because the explicit debris term is omitted from these residual definitions,
the normalized Poisson equation requires both residuals to recover
$\rho_{d}$, apart from deposition and differentiation errors. This
is a verification of the field reconstruction and source deposition,
not a remote-localization result. The configured $x_{d}$ is not used
in locating the residual maximum. In Table. \ref{tab:Poisson-residual-localization.},
we list the physical parameters of the simulation. In the table, the
flow velocity is written in terms of the Mach number $M$. The configured
debris position is $x_{d}$ in terms of Debye length $\lambda_{De}$,
and $\widehat{x}_{d}$ is the discovered position, estimated from
the residuals. For the case presented in Table. \ref{tab:Poisson-residual-localization.},
we have the mean absolute error (MAE) and median, where $N$ is the
number of flow cases, 
\begin{eqnarray}
\mathrm{MAE} & = & \frac{1}{N}\sum^{N}_{j=1}\left|\widehat{x}_{d,j}-x_{d,j}\right|=0.441\lambda_{De},\\
\mathrm{median}(|\widehat{x}_{d}-x_{d}|) & = & 0.340\lambda_{De},\\
\max|\widehat{x}_{d}-x_{d}| & = & 1.110\lambda_{De}.
\end{eqnarray}
In Fig. \ref{fig:Configured-debris-position}, we compare the configured
position $x_{d}$ with the position $\widehat{x}_{d}$ obtained from
the residual maximum. The error remains below $1.5\lambda_{De}$ for
all flow cases. This provides a numerical benchmark for the subsequent
analysis, which depends on accurate reconstruction of the plasma fields.

\begin{figure}[t]
\begin{centering}
\includegraphics[width=1\textwidth]{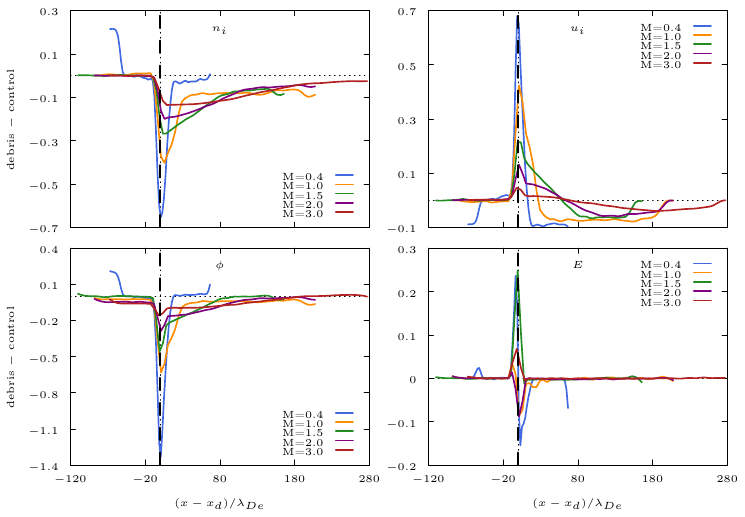} 
\par\end{centering}
\caption{Time-averaged differences between the debris and control profiles
for several plasma fields, plotted relative to the debris position.}
\label{fig:Time-averaged-difference-between} 
\end{figure}

\subsection{The diagnostic regions and dispersion power spectra}

\begin{figure}
\begin{centering}
\includegraphics[width=0.5\textwidth]{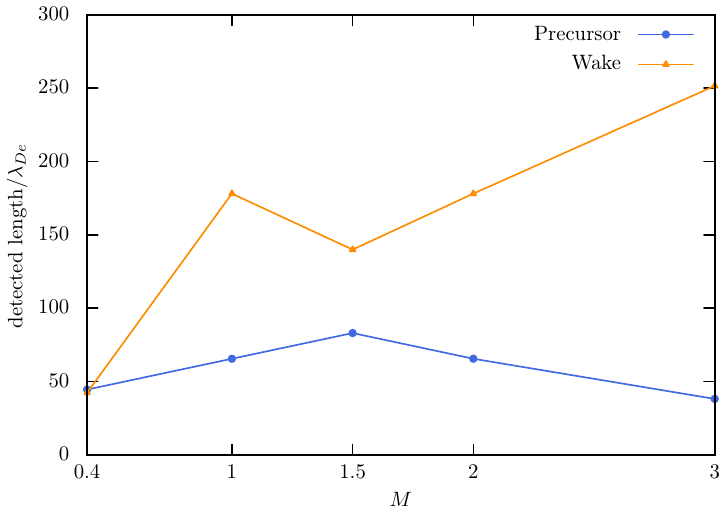} 
\par\end{centering}
\caption{The extent and variation of the precursor and wake regions with Mach
number $M$.}
\label{fig:The-extent-and} 
\end{figure}

\begin{figure}
\begin{centering}
\includegraphics[width=1\columnwidth]{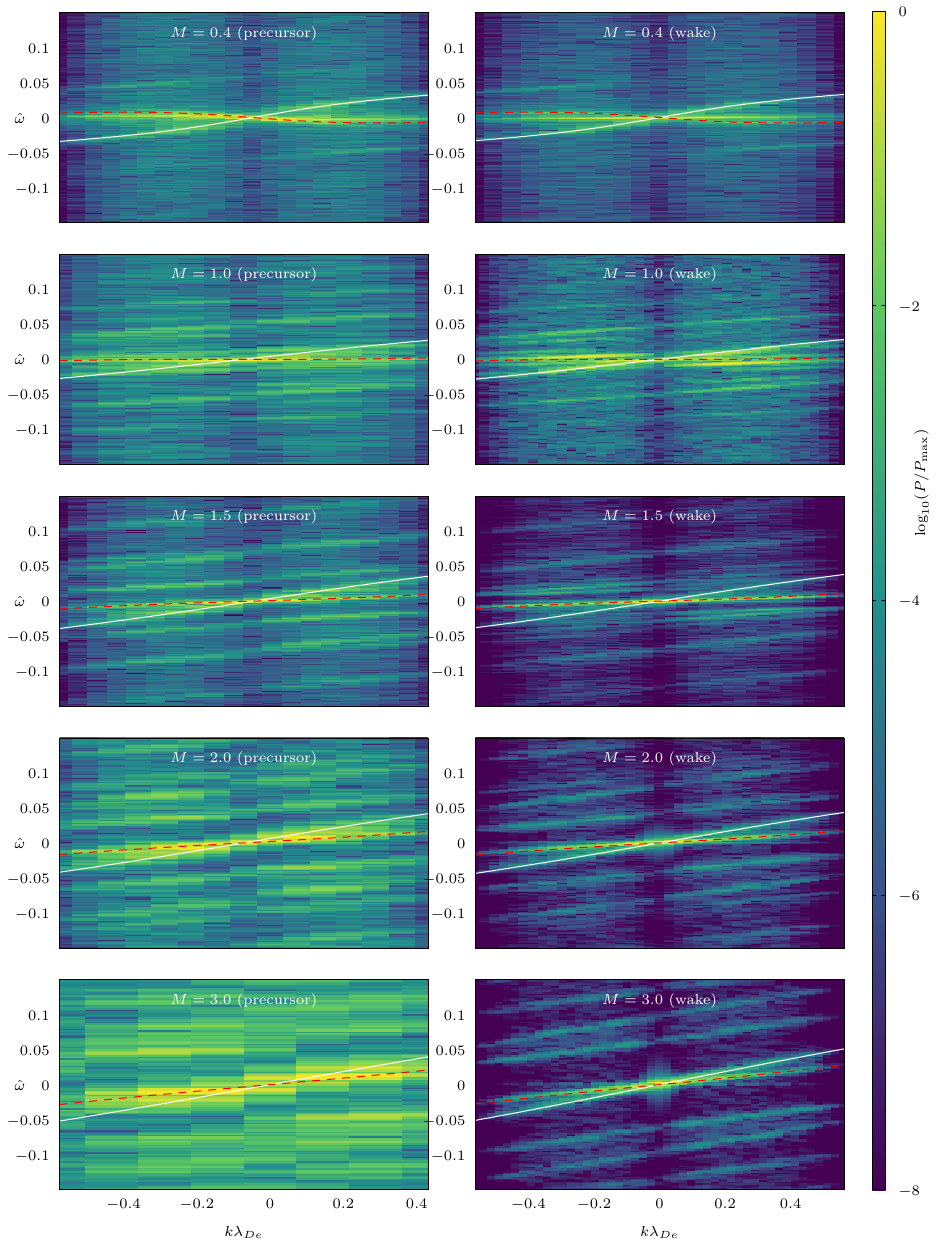} 
\par\end{centering}
\caption{The $\omega$-$k$ spectra for the precursor and wake regions for
different $M$. The dispersion branches $\omega_{+}$ (white solid)
and $\omega_{-}$ (red dashed) are superimposed on the heat maps.
The plotted frequency is $\widehat{\omega}=\omega_{\mathrm{phys}}/\omega_{\textrm{pi}}=(\omega_{\mathrm{phys}}/\omega_{\textrm{pe}})\sqrt{m_{i}/m_{e}}$.}
\label{fig:The---spectra} 
\end{figure}

The auto-detection of the diagnostic regions is shown in Fig. \ref{fig:Time-averaged-difference-between}
and the extent of the precursor and wake regions are shown in Fig.
\ref{fig:The-extent-and}. The dispersion ($\omega$-$k$) power spectra
of the precursor and wake regions for various $M$ are shown in Fig.
\ref{fig:The---spectra}. As we have pointed out in Eq. (\ref{eq:debris-fluct}),
the composite disturbance of field $\delta g_{d}(x,t)$ due to the
debris can be written as 
\begin{equation}
\delta_{t}g_{d}(x,t)=g_{d}(x,t)-\langle g_{d}(x,t)\rangle_{t}.
\end{equation}
The $\omega$-$k$ power spectrum is then calculated from the double
Fourier transform as 
\begin{equation}
P(k,\omega)=\left|\widehat{\delta_{t}g_{d}}(k,\omega)\right|^{2},
\end{equation}
where 
\begin{equation}
\widehat{\delta_{t}g_{d}}(k,\omega)=\int_{\mathcal{T}}\int_{\mathcal{R}}\delta_{t}g_{d}(x,t)e^{-i(kx-\omega t)}\,dx\,dt.
\end{equation}
Here ${\cal R}$ is the selected spatial diagnostic region and ${\cal T}$
is the selected time interval. The plotted heat-map is approximately
as per 
\begin{equation}
\log_{10}\left[\frac{P(k,\omega)}{P_{\max}}\right].
\end{equation}
The inclined ridges in almost all the panels approximately satisfy
\begin{equation}
\omega\simeq v_{\mathrm{ph}}k,
\end{equation}
where $v_{\mathrm{ph}}$ is the phase velocity of the underlying wave,
in the same normalized velocity units as $x/t$. The fact that the
supersonic flows $(M>1)$ especially show these ridges much more clearly,
indicates that the high-flow wake contains organized propagating electrostatic
structures. As most power lies at $\omega\ll\omega_{\textrm{pe}}$,
electron plasma oscillations do not play any significant part in this
instability, which is also supported by the electron distribution
function. Also much of the power occurs at relatively small $|k\lambda_{De}|$,
indicating structures extending over several Debye lengths. Multiple
branches are observed at high $M$, the $M=3$ wake contains several
parallel ridges. In all possibilities, these may be due to multiple
phase velocities, harmonics, nonlinear sidebands, or several convected
wake structures. The near-zero frequency component (power close to
$\omega=0$) represents slowly evolving or quasi-stationary wake and
sheath structures.

\subsubsection{Ion-acoustic (IA) branches}

The normalized dispersion relation for the warm IA can be written
as 
\begin{equation}
\omega^{2}_{\textrm{IA}}=k^{2}\left(\frac{1}{1+k^{2}}+\gamma_{i}\frac{T_{i}}{T_{e}}\right),
\end{equation}
where $\gamma_{i}\equiv3$ is the ratio of specific heats of ions
in 1D. In a flowing plasma, however, we have two Doppler-shifted branches
\begin{equation}
\omega_{\pm}=Mk\pm\omega_{\textrm{IA}}.
\end{equation}
For sufficiently small $k$, we can approximate the above relation
as 
\begin{equation}
\omega_{\pm}\simeq k(M\pm1)
\end{equation}
in the cold ion limit. In the above expressions, $\omega$ is normalized
by the ion-plasma frequency $\omega_{\textrm{pi}}$, $k$ by $\lambda^{-1}_{De}$.
In Fig. \ref{fig:The---spectra}, these branches are superimposed
as a white solid line ($\omega_{+}$) and a red dashed line ($\omega_{-}$).
Several prominent parallel ridges appearing in high $M$ wake regions
may indicate modulation or sidebands of the IA disturbance, coherent
wave packets, other harmonics generated through nonlinear mechanisms,
or genuinely coupled nonlinear modes.

\begin{figure}[t]
\begin{centering}
\includegraphics[width=1\columnwidth]{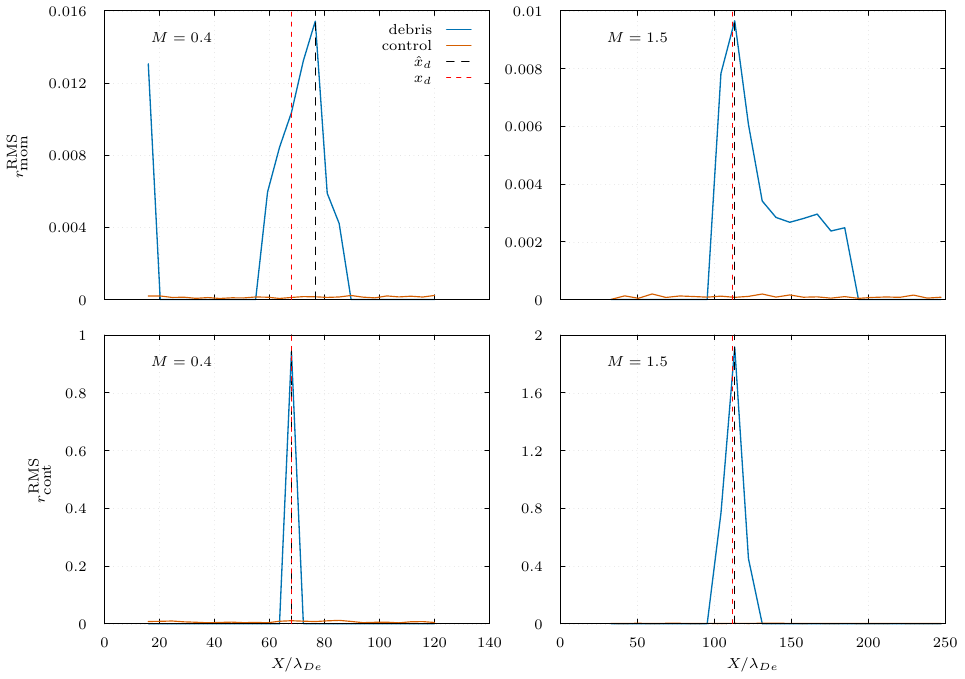} 
\par\end{centering}
\caption{Fluid-equation residuals for representative subsonic and supersonic
flows. The excess over the matched control identifies the localized
departure from the bulk equations.}
\label{fig:A-verification-of} 
\end{figure}

\subsection{Control-trained fluid residuals}

Weak-form regression is first applied to the control simulation to
recover the coefficients of the expected ion continuity and momentum
balances. The purpose is not to rediscover standard fluid theory,
but to establish a data-consistent bulk model against which the debris
run can be compared.

The equations to be discovered are 
\begin{eqnarray}
\partial_{t}n_{i} & = & -\xi_{c}\partial_{x}(n_{i}u_{i}),\\
\partial_{t}u_{i} & = & -\xi_{m,1}u_{i}\partial_{x}u_{i}-\xi_{m,2}\frac{q_{i}}{m_{i}}\partial_{x}\phi-\xi_{m,3}\frac{1}{m_{i}n_{i}}\partial_{x}p_{i},
\end{eqnarray}
respectively for the ion continuity and ion momentum equations, where
$\xi_{c}$ and $\bm{\xi}_{m}$ are fitted from the control PIC data.
For example, the weak-form estimates the coefficients for the control
run for plasma flow $M=0.4$ and $1.5$ are 
\begin{eqnarray}
M=0.4 & \Rightarrow & \left\{ \begin{array}{rcl}
\xi_{c} & = & 1.0057,\\
\bm{\xi}_{m} & = & (1.0516,0.99,1.0932).
\end{array}\right.\\
M=1.5 & \Rightarrow & \left\{ \begin{array}{rcl}
\xi_{c} & = & 1.0509,\\
\bm{\xi}_{m} & = & (1.0878,0.95,1.0).
\end{array}\right.
\end{eqnarray}
The ideal value of each coefficient is unity, and the recovered values
are close to this limit. In the debris run, however, particle collection
and the localized plasma response introduce an effective source term,
\begin{equation}
\partial_{t}n_{i}+\partial_{x}(n_{i}u_{i})=S_{n,d}(x,t),
\end{equation}
where $S_{n,d}(x,t)$ represents the debris-induced contribution.
The corresponding control-trained continuity residual is 
\begin{equation}
r_{c}=\partial_{t}n_{i,d}+\xi_{c}\partial_{x}(n_{i,d}u_{i,d}),
\end{equation}
where $(n_{i,d},u_{i,d})$ are the density and plasma velocity determined
from the moments of the ion distribution function in presence of the
debris with SINDy coefficient $\xi_{c}$ determined from the control
run.

\begin{figure}[t]
\begin{centering}
\includegraphics[width=1\columnwidth]{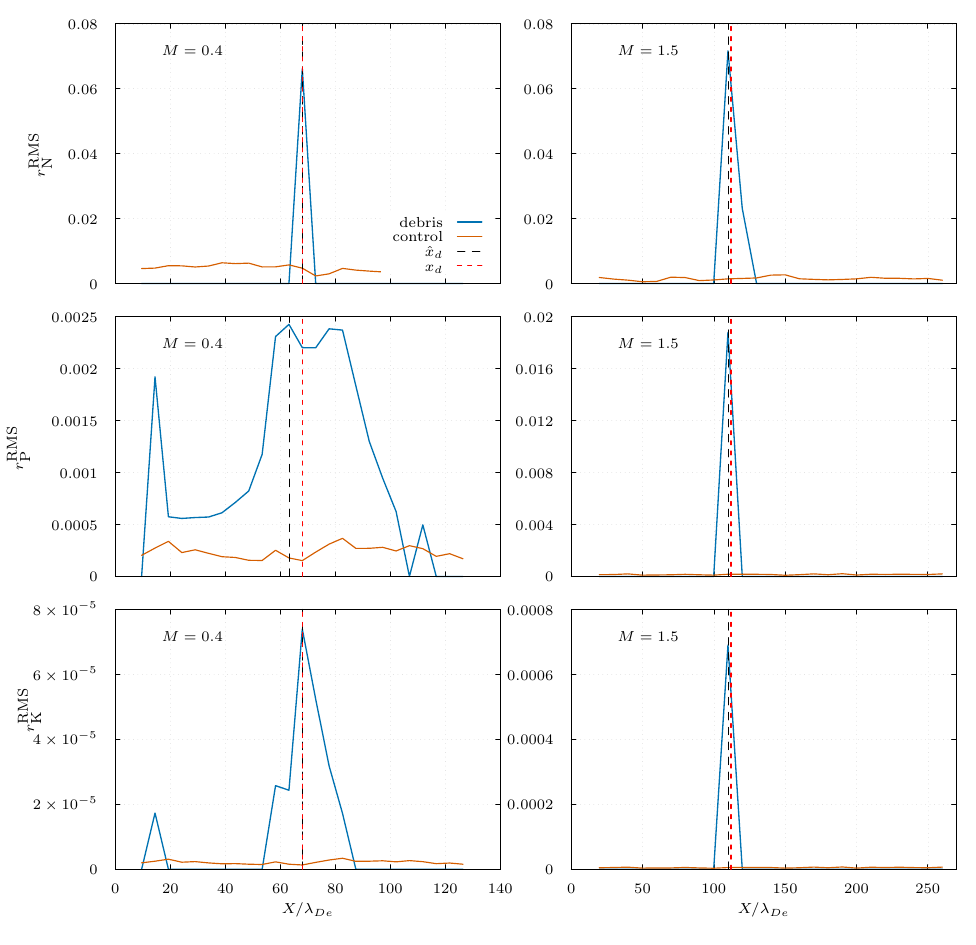} 
\par\end{centering}
\caption{Velocity moments of the control-trained Vlasov residual for representative
subsonic and supersonic flows.}
\label{fig:A-verification-of-1} 
\end{figure}

The position of the largest localized departure is estimated from
the statistically largest residual excess, 
\begin{equation}
\widehat{x}_{d}=x_{\ell_{*}},\qquad\ell_{*}=\operatorname*{arg\,max}_{\ell}Z_{\ell}.
\end{equation}
The results for the localization of the debris based continuity and
momentum residual for two representative subsonic and supersonic plasma
flows with $M=0.4$ and $1.5$ are shown in Fig. \ref{fig:A-verification-of}.
$\sigma_{\mathrm{control}}$ is the temporal standard deviation in
the corresponding no-debris control run. For a diagnostic quantity
$g(x,t)$, by Eq. (\ref{eq:debris-fluct}), the control fluctuation
can be written as
\begin{equation}
\delta_{t}g_{\mathrm{control}}(x,t)=g_{\mathrm{control}}(x,t)-\left\langle g_{\mathrm{control}}(x,t)\right\rangle _{t},
\end{equation}
and its temporal standard deviation is given by
\begin{equation}
\sigma_{\mathrm{control}}(x)=\sqrt{\left\langle \left[\delta_{t}g_{\mathrm{control}}(x,t)\right]^{2}\right\rangle _{t}}.
\end{equation}
In Fig. \ref{fig:A-verification-of}, we plot $r^{\textrm{RMS}}_{\textrm{cont},\textrm{mom}}$,
the RMS values of the residual $r_{\textrm{cont},\textrm{mom}}$ for
continuity and momentum, calculated over $N$ different sampling points
within the neighborhood $X$, 
\begin{equation}
r^{\textrm{RMS}}(X)=\sqrt{\frac{1}{N}\sum^{N}_{j=1}r^{2}_{j}}
\end{equation}
against the neighborhood $X$. The localization with Vlasov residuals
are shown in Fig. \ref{fig:A-verification-of-1}.

\subsection{Localization from remote disturbance envelopes}

We now examine how the debris-induced disturbance varies with distance
from its source. An empirical spatial profile is learned from the
training cases and translated across a trial domain. Sparse regression
is used only to select a compact dependence of the disturbance envelope
on equilibrium flow, debris charge, and source distance.

\subsubsection{Disturbance envelope}

For each plasma field 
\begin{equation}
g\in\{n_{i},u_{i},\phi,E\},
\end{equation}
we find out the temporal means $(\mu)$ and variances $(V)$ for both
the control run and debris run 
\begin{equation}
\mu_{g,d}(x),\quad V_{g,d}(x),\quad\mu_{g,c}(x),\quad V_{g,c}(x),
\end{equation}
where the subscripts $(c,d)$ denote the control and debris runs.
The temporal mean is as defined by Eqs. (\ref{eq:temporal_mean-1})
and (\ref{eq:temporal_mean-2}) 
\begin{equation}
\mu_{g}(x)=\left\langle g(x,t)\right\rangle _{t}
\end{equation}
and the variance is given by 
\begin{equation}
V_{g}(x)=\left\langle \left[g(x,t)-\mu_{g}(x)\right]^{2}\right\rangle _{t}.
\end{equation}
The disturbance envelope $D_{g}(x)$ for a particular plasma field
$g$ is defined as 
\begin{equation}
D_{g}(x)=\sqrt{\left[\mu_{g,d}(x)-\mu_{g,c}(x)\right]^{2}+\max\left[V_{g,d}(x)-V_{g,c}(x),0\right]}.
\end{equation}
The disturbance envelope, $D_{g}$ contains both the stationary debris-induced
profile and the excess temporal fluctuation above the control run.
The noise threshold of the disturbance is defined as 
\begin{equation}
\tilde{\sigma}_{g,c}(x)=\sqrt{\frac{V_{g,c}(x)}{N_{c}}},
\end{equation}
where $N_{c}$ is the number of selected control snapshots. Only points
satisfying 
\begin{equation}
D_{g}(x)>\tilde{\sigma}_{g,c}(x)
\end{equation}
are retained as valid debris-induced disturbance.

\subsubsection{Empirical envelope model and training}

For a training case with source position $x_{d}$, we define the distance
$s$, 
\begin{equation}
r_{d}=x-x_{d},\qquad s=|r_{d}|.
\end{equation}
Only observations outside the debris cell and its immediate sheath
are used: 
\begin{equation}
8\lambda_{De}\leq s\leq70\lambda_{De}
\end{equation}
Thus, the empirical model is fitted only to remote precursor and wake
fields, not to measurements from the source region. Source positions
for the training cases are supplied by the preceding Poisson-residual
benchmark, whereas the position of the withheld trial case is not
supplied to the envelope model.

We note that the electrostatic disturbance produced by the debris
is analogous to the phenomenon of Debye shielding and the disturbance
$D_{g}(s)$ is expected to follow the same expression as the electrostatic
potential $\phi$ in a Debye shielding case 
\begin{equation}
\frac{\partial D_{g}}{\partial s}=-\kappa_{g}(M,\widehat{Q}_{d})D_{g},
\end{equation}
where $\kappa_{g}$ is an effective inverse attenuation length and
where $\widehat{Q}_{d}$ is the normalized debris charge 
\begin{equation}
\widehat{Q}_{d}=\frac{Q_{d}/e}{10^{9}}.
\end{equation}
Note that $Q_{d}$ is connected to the charge density through the
relation 
\begin{equation}
Q_{d}(t)=\int^{L}_{0}\rho_{d}(x,t)\,dx,
\end{equation}
where $L$ is the extent of the Gaussian charge profile satisfying
\begin{equation}
\int^{L}_{0}G_{d}(x)\,dx=1.
\end{equation}
This motivates the compact candidate model 
\begin{equation}
\frac{\partial D_{g}}{\partial s}=\left(\xi_{0,g}+\xi_{M,g}M+\xi_{Q,g}\widehat{Q}_{d}\right)D_{g}\equiv\Gamma_{g,\alpha}(M,\widehat{Q}_{d})D_{g},
\end{equation}
where 
\begin{equation}
\Gamma_{g,\alpha}(M,\widehat{Q}_{d})=\xi_{0,g,\alpha}+\xi_{M,g,\alpha}M+\xi_{Q,g,\alpha}\widehat{Q}_{d}
\end{equation}
where the coefficients are estimated by sparse regression. For constant
$M$ and approximately stationary debris charge, integration gives
\begin{equation}
D_{g}(s)=D_{0,g}(M,\widehat{Q}_{d})\,e^{\Gamma_{g}(M,\widehat{Q}_{d})s}
\end{equation}
or equivalently 
\begin{equation}
\log D_{g}(s)=\log D_{0,g}+s\left(\xi_{0,g}+\xi_{M,g}M+\xi_{Q,g}\widehat{Q}_{d}\right).
\end{equation}
The source amplitude may also depend on flow and debris charge with
\begin{equation}
\log D_{0,g}=b_{0,g}+b_{M,g}M+b_{Q,g}\widehat{Q}_{d}.
\end{equation}
Combining the two relations gives the empirical model 
\begin{equation}
\log D_{g}(s)=b_{0}+b_{M}M+b_{Q}\widehat{Q}_{d}+s\left(\gamma_{0}+\gamma_{M}M+\gamma_{Q}\widehat{Q}_{d}\right)
\end{equation}

\begin{figure}[t]
\begin{centering}
\includegraphics[width=1\columnwidth]{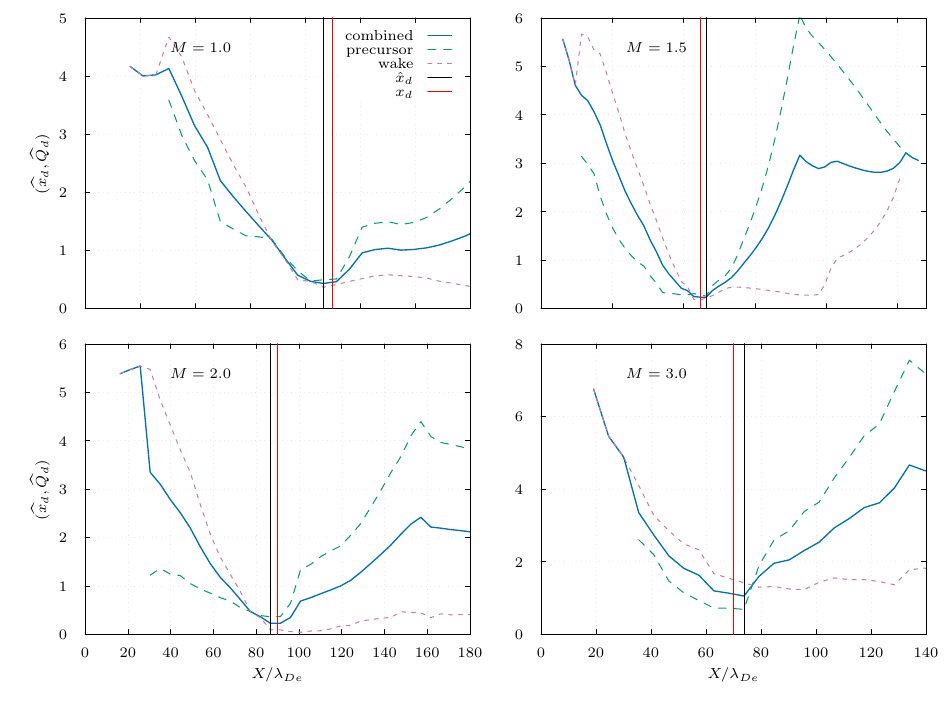} 
\par\end{centering}
\caption{Localization obtained by translating the empirical precursor and wake
envelopes across the trial domain. The vertical lines mark the configured
and estimated source positions. The method resolves the transonic
and supersonic cases but not the subsonic cases.}
\label{fig:The-remote-and} 
\end{figure}

A compact regression representation is 
\begin{equation}
y_{g,\alpha}=\boldsymbol{\Psi}\boldsymbol{\xi}_{g,\alpha},\qquad y_{g,\alpha}=\log D_{g,\alpha},
\end{equation}
where coefficient vector is given by 
\begin{equation}
\boldsymbol{\xi}_{g,\alpha}=\left[b_{0},\,b_{M},\,b_{Q},\,\gamma_{0},\,\gamma_{M},\,\gamma_{Q}\right]^{\mathsf{T}}_{g,\alpha}.
\end{equation}
and the candidate library is 
\begin{equation}
\boldsymbol{\Psi}=\left[1,\,M,\,\widehat{Q}_{d},\,s,\,sM,\,s\widehat{Q}_{d}\right].
\end{equation}
In the above expressions, we have used $\alpha$ to denote either
of the precursor or wake regions. The sparse regression therefore
gives an equation of the form 
\begin{equation}
\log D_{g,\alpha}(s)=b_{0,\alpha}+b_{M,\alpha}M+b_{Q,\alpha}\widehat{Q}_{d}+s\left(\gamma_{0,\alpha}+\gamma_{M,\alpha}M+\gamma_{Q,\alpha}\widehat{Q}_{d}\right)
\end{equation}
Equivalently, the empirical relation is 
\begin{equation}
D_{g,\alpha}(s)=D_{0,g,\alpha}\left(M,\widehat{Q}_{d}\right)e^{\Gamma_{g,\alpha}\left(M,\widehat{Q}_{d}\right)s},
\end{equation}
where 
\begin{eqnarray}
D_{0,g,\alpha} & = & \exp\left(b_{0,\alpha}+b_{M,\alpha}M+b_{Q,\alpha}\widehat{Q}_{d}\right),\\
\Gamma_{g,\alpha} & = & \gamma_{0,\alpha}+\gamma_{M,\alpha}M+\gamma_{Q,\alpha}\widehat{Q}_{d}.
\end{eqnarray}
This model describes how the disturbance amplitude and its effective
spatial growth or attenuation rate depend on $M$ and $Q_{d}$, separately
in the precursor and wake regions.

In this simulation, we use six equilibrium-flow cases,
\begin{equation}
M=0,0.4,1.0,1.5,2.0,3.0.
\end{equation}
One case at a time is withheld as a trial, while the other five cases
are used for training.

\subsubsection{Blind localization of the debris}

Each simulation is successively treated as an independent trial. The
position used to generate that case is withheld from the empirical
model, which is constructed using the other five simulations. The
Mach number is treated as a measured background parameter, while the
source position and charge are scanned.

For every measurement position $x_{j}$, we find 
\begin{equation}
s_{j}(X)=|x_{j}-X|,
\end{equation}
where $X$ is the trial source position. The trial Mach number $M$
is known, while $Q_{d}$ is scanned over the range found in the training
cases. For every field and side, the model predicts $D^{\mathrm{pred}}_{g,\alpha}$
as against the observed disturbance $D^{\mathrm{obs}}_{g,\alpha}$,
\begin{equation}
\log D^{\mathrm{pred}}_{g,\alpha}\left[s_{j}(X);M,Q_{d}\right],
\end{equation}
while the model mismatch is given by 
\begin{equation}
J_{g,\alpha}(X,Q_{d})=\frac{1}{N_{g,\alpha}}\sum_{j\in\alpha}\left[\log D^{\mathrm{obs}}_{g,\alpha}(x_{j})-\log D^{\mathrm{pred}}_{g,\alpha}(s_{j})\right]^{2}.
\end{equation}
The model mismatches are then averaged separately for both precursor
$(\alpha=p)$ and wake $(\alpha=w)$, 
\begin{eqnarray}
J_{p}(X,Q_{d}) & = & \left\langle J_{g,p}(X,Q_{d})\right\rangle _{g},\\
J_{w}(X,Q_{d}) & = & \left\langle J_{g,w}(X,Q_{d})\right\rangle _{g}.
\end{eqnarray}
The combined mismatch is 
\begin{equation}
J(X,Q_{d})=\frac{J_{p}(X,Q_{d})+J_{w}(X,Q_{d})}{2},
\end{equation}
Equal precursor--wake weighting prevents a long supersonic wake from
numerically overwhelming the precursor information. Finally, the quantity
\begin{equation}
(\widehat{x}_{d},\widehat{Q}_{d})=\underset{X,Q_{d}}{\operatorname{arg\,min}}\,J(X,Q_{d})
\end{equation}
provides the estimated source position and charge for the trial case,
as plotted in Fig. \ref{fig:The-remote-and}.

For the transonic and supersonic cases, the estimated position agrees
with the configured value with a mean absolute error of approximately
$3.58\lambda_{De}$. The subsonic cases are not resolved. This limitation
is consistent with their weaker and less spatially differentiated
precursor--wake signatures and defines the present range of validity
of the approach.

\section{Summary and scope}\label{sec:Summary-and-scope}

We have examined whether the nonlinear plasma response to a localized
charged object retains information about the position of its source.
The open-boundary PIC calculation produces a sustained counter-streaming
kinetic state together with spatially asymmetric precursor and wake
disturbances. A matched debris-free calculation separates these disturbances
from the background fluctuations. Their spatial extent and frequency--wavenumber
spectra show that the response remains organized even after a reduced
fKdV description ceases to reproduce the fully developed kinetic state.

Weak-form fluid and Vlasov residuals provide compact diagnostics of
the debris-induced departure from the control dynamics. More importantly
for the inverse problem, disturbance envelopes sampled outside the
debris cell and its immediate sheath can be represented by a simple
flow and charge-dependent spatial model. Its predictive capability
is examined by successively treating each simulation as an independent
case and constructing the model from the remaining simulations. Translating
the resulting profile across the trial domain recovers the source
position for transonic and supersonic flows, with a mean absolute
error of approximately $3.58\lambda_{De}$. The failure of the subsonic
cases is equally informative: within the present model, their precursor
and wake do not provide a sufficiently distinct spatial signature
for localization.

These results establish a possible plasma-based route to source localization,
not a ready-to-implement debris-detection procedure. The present conclusions
are restricted to a one-dimensional electrostatic, cold-ion model
and to the parameter ensemble examined here. Extension to multidimensional
geometry, mixed $\textrm{H}^{+}$--$\textrm{O}^{+}$ plasmas, magnetic
fields, photoemission, broader background variability, and independent
noise realizations is required before the relevance to a particular
LEO environment can be assessed quantitatively. The contribution of
the present work is the demonstration that the remote plasma response
can possess a reproducible, source-centered structure from which positional
information may be inferred.

\subsection{Prognosis}

The present results suggest a possible route toward a more general
data-driven localization framework. A multidimensional PIC model could
generate libraries of precursor and wake responses over a broad range
of plasma composition, temperature, density, magnetic field, object
charge, relative velocity, and observation geometry. Such synthetic
data could subsequently be used to train an artificial neural network
to infer the probable position and other parameters of a charged object
from plasma measurements acquired away from its immediate sheath.

This possibility should presently be regarded as a prognosis rather
than a proposed detection system. The one-dimensional results establish
only that the debris-induced plasma response can possess a reproducible,
source-centered structure and that this structure contains sufficient
information for localization in the transonic and supersonic cases
examined here. Whether an artificial neural network can retain this
capability under realistic multidimensional LEO conditions, natural
plasma variability, measurement noise, and incomplete spatial sampling
remains to be investigated.

\section*{Acknowledgement}

The authors gratefully acknowledge D.\ Chakrabarty of the Physical
Research Laboratory, Ahmedabad, and U.\ Kumar of the Indian Space
Research Organisation (ISRO), Bengaluru, for their constructive inputs
during the course of this work. This work was supported by ISRO’s
RESPOND Programme under Project Grant No. RAC-S/GU/2024/4/74.\vfill{}

\emph{\pagebreak}

\section*{Appendix}

\subsection*{A. Baseline PIC simulation parameters}

Table. \ref{tab:pic-base-parameters} lists the parameters common
to the open-boundary PIC calculations. Run-dependent quantities, particularly
the Mach number, debris position, number of time steps, final time,
and selected analysis interval, are deliberately omitted and are stated
with the corresponding runs in the main text.

\begin{table}[H]
\centering{}\caption{Baseline parameters for the open-boundary PIC simulations. Lengths,
times, velocities, densities, and potential are normalized by $\lambda_{De}$,
$\omega^{-1}_{\textrm{pe}}$, $v_{\textrm{th}e0}$, $n_{0}$, and
$T_{e0}/e$, respectively.\vrd}\label{tab:pic-base-parameters}
\begin{tabular}{lll}
\hline 
\textbf{Category}\vrr & \textbf{Parameter} & \textbf{Value}\tabularnewline
\hline 
Model\vrt & Geometry and field model & One-dimensional electrostatic PIC\tabularnewline
 & Boundary condition & Open reservoirs with two-sided thermal injection\tabularnewline
 & Domain length $L$ & $136\lambda_{De}$ $^{\dag}$\tabularnewline
 & Grid points $N_{x}$ & $1024$ $^{\dag}$\tabularnewline
 & Grid spacing $\Delta x$ & $0.133\lambda_{De}$\tabularnewline
 & Time step $\omega_{\textrm{pe}}\Delta t$ & $0.05$\vrd\tabularnewline
\hline 
Plasma loading\vrt & Electron/ion macro-particles & $167000$ per species$^{\ddag}$\tabularnewline
 & Initial electron/ion density & $n_{e}=n_{i}=1$\tabularnewline
 & Electron/ion temperature & $T_{e}=0.5$, $T_{i}=0.1$ (normalized)\tabularnewline
 & Ion-to-electron mass ratio & $m_{i}/m_{e}=1836$\vrd\tabularnewline
\hline 
Numerical method\vrt & Charge deposition/interpolation & Cloud-in-cell\tabularnewline
 & Particle advance & Leapfrog\tabularnewline
 & Field solution & Dirichlet Poisson solver\tabularnewline
 & Boundary potentials & $\phi(0)=\phi(L)=0$\vrd\tabularnewline
\hline 
Reservoir buffer\vrt & Width & $20\lambda_{De}$ at each boundary\tabularnewline
 & Maximum relaxation rate & $\alpha_{\max}=0.20$\tabularnewline
 & Relaxation exponent & $\kappa=2$\tabularnewline
 & Reset probability & $1-\exp[-\alpha(x)\Delta t]$\vrd\tabularnewline
\hline 
Debris model\vrt & Collection radius $R_{d}$ & $0.05\lambda_{De}$\tabularnewline
 & Charge representation & Truncated Gaussian\tabularnewline
 & Gaussian width & $\sigma_{d}=R_{d}/3$\tabularnewline
 & Initial charge and charging & $Q_{d}(0)=0$; dynamic charging enabled\tabularnewline
 & Physical particles per macro-particle & $10^{6}$\tabularnewline
 & Maximum capture probabilities & $P_{\max,e}=0.010$, $P_{\max,i}=0.005$\tabularnewline
 & Capture-energy scales & $E_{\mathrm{scale},e}=0.20$, $E_{\mathrm{scale},i}=0.05$\tabularnewline
 & Threshold energy & $E_{\mathrm{th},e}=E_{\mathrm{th},i}=0$\tabularnewline
 & Photoemission & Disabled\vrd\tabularnewline
\hline 
Diagnostics\vrt & Vlasov snapshot cadence & Every $50$ steps $\left(\Delta t_{\mathrm{snap}}=2.5\omega^{-1}_{\textrm{pe}}\right)$\tabularnewline
 & Phase-space histogram & $64\times64$ bins\tabularnewline
 & Electron/ion velocity ranges & $[-5,5]v_{\textrm{th}e}$, $[-0.08,0.08]v_{\textrm{th}e}$\tabularnewline
 & Probe half-widths & $1\lambda_{De}$ (local), $4\lambda_{De}$ (instability)\tabularnewline
 & Spatial-derivative smoothing & Two binomial passes\vrd\tabularnewline
\hline 
\end{tabular}
\end{table}

$^{\dag}$\emph{Can be arbitrarily increased or decreased, keeping
the prescribed plasma parameters same.}

$^{\ddag}$\emph{Will scale automatically depending on what plasma
parameters are prescribed.}

\vfill{}

\pagebreak{}

\subsection*{B. Mills-ratio expansion}

For $x>0$, let's define the upper Gaussian tail 
\begin{equation}
Q(x)=\Theta(-x)=\int^{\infty}_{x}\psi(t)\,dt.
\end{equation}
The Mills ratio is given by 
\begin{equation}
R(x)=\frac{Q(x)}{\psi(x)}=\frac{\Theta(-x)}{\psi(x)}
\end{equation}
and subsequently 
\begin{equation}
\Theta(-x)=\psi(x)\,R(x).
\end{equation}
For $|x|\gg1$, the Mills-ratio asymptotic expansion for a standard
normal distribution is given by \citep{Baricz2008} 
\begin{equation}
R(x)\sim\sum^{\infty}_{k=0}(-1)^{k}\frac{1\cdot3\cdot5\cdots(2k-1)}{x^{2k+1}},
\end{equation}
or explicitly 
\begin{equation}
R(x)\sim\frac{1}{x}-\frac{1}{x^{3}}+\frac{3}{x^{5}}-\frac{15}{x^{7}}+\frac{105}{x^{9}}-\cdots.\label{eq:mills-expansion}
\end{equation}
for large positive $x$ and this expansion is used to calculate the
tail probability in our case for large streaming velocity.

For the case represented by Eq. (\ref{eq:mills-flux}), we have for
the right flux 
\begin{equation}
\Gamma_{s,\textrm{right}}=n_{s}v_{\textrm{th},s}\psi(a_{s})\left[1-a_{s}R(a_{s})\right].
\end{equation}
Using Eq. (\ref{eq:mills-expansion}), the counter-flowing flux is
consequently evaluated as 
\begin{equation}
1-a_{s}R(a_{s})\sim\frac{1}{a^{2}_{s}}-\frac{3}{a^{4}_{s}}+\frac{15}{a^{6}_{s}}-\frac{105}{a^{8}_{s}}+\cdots
\end{equation}
instead of directly subtracting two nearly equal quantities, which
is numerically much more reliable. At present, we retain terms of
the above expansion up to the eighth order.

\vfill{}
\pagebreak{}

\subsection*{C. Regression and localization parameters}

\begin{table}[H]
\centering{}\caption{Parameters used for weak regression, residual construction, and remote-envelope
localization. Spatial widths are expressed in electron Debye lengths
and temporal windows in saved snapshots.\vrd}\label{tab:analysis-parameters}
\begin{tabular}{ll}
\hline 
\textbf{Parameter}\vrr & \textbf{Value}\tabularnewline
\hline 
Debris/control analysis interval\vrt & $\omega_{\mathrm{pe}}t\geq10^{4}$ / full interval\tabularnewline
General integral window/stride & $400/40$\tabularnewline
Weak spatial half-width $h_{x}$ & $3$ cells\tabularnewline
Localized weak window/stride & $200/20$\tabularnewline
STLSQ threshold scan & $24$ logarithmic values, $10^{-5}$--$10^{-0.5}$\tabularnewline
Ridge parameter & $10^{-8}$\tabularnewline
Maximum STLSQ iterations & $25$\tabularnewline
Validation fraction & $0.25$\tabularnewline
Complexity penalty & $0.002$\tabularnewline
Fluid stencil $(h_{x},N_{t},\Delta N_{t})$ and trim & $(2,9,8)$, $4$ bins\tabularnewline
Kinetic stencil $(h_{x},h_{v},\Delta v,N_{t},\Delta N_{t})$ & $(2,3,3,9,16)$\tabularnewline
Gaussian trial centers and width & $25$, $\sigma_{G}=3\lambda_{De}$\tabularnewline
Source-fit ridge parameter & $10^{-8}$\tabularnewline
Relative group threshold/iterations & $0.20/8$\tabularnewline
Residual acceptance thresholds & $Z_{\mathrm{thr}}=5$, $S_{\mathrm{thr}}=2$\tabularnewline
Position-agreement tolerance & $6\lambda_{De}$\tabularnewline
Vlasov species and smoothing & Ions, one pass\tabularnewline
Vlasov $(x,v)$ trim & $(4,2)$ bins\tabularnewline
Vlasov residual window/stride & $40/40$\tabularnewline
Velocity-scan step and extent & $0.25c_{s}$, at least $\pm2c_{s}$\tabularnewline
Remote mask and fit radius & $8\lambda_{De}$, $70\lambda_{De}$\tabularnewline
Remote source width and radius & $4\lambda_{De}$, $30\lambda_{De}$\tabularnewline
Charge scale & $10^{9}e$\tabularnewline
Inverse threshold scan & $18$ logarithmic values, $10^{-3}$--$10^{-0.3}$\tabularnewline
Maximum relative model RMSE\vrd & $0.90$\tabularnewline
\hline 
\end{tabular}
\end{table}
\vfill{}
\pagebreak{}

 \bibliographystyle{plain}
\bibliography{ref}

\end{document}